\documentclass[a4paper,11pt]{article}
\usepackage{jcappub}
\usepackage{lineno}

\usepackage{amssymb}
\usepackage{ulem}

\usepackage{color}
\usepackage{gensymb}
\usepackage{siunitx}
\usepackage{MnSymbol}
\usepackage{subcaption}
\usepackage{xcolor}
\usepackage{tabularx}
\usepackage{array}
\newcolumntype{C}[1]{%
    >{\hsize=#1\hsize\linewidth=\hsize
      \centering\arraybackslash}X%
}
\usepackage{multirow}
\usepackage{chngcntr}

\usepackage{tabularx}

\usepackage{mathtools}
\usepackage{verbatim}
\usepackage{graphicx} 
\usepackage{subcaption}
\usepackage{url}

\usepackage{placeins}

\title{\boldmath Low-Mass Magnetic Monopoles in the Galaxy: Simulations and Comparison with Ultra-High-Energy Cosmic-Ray Data}

\author[a]{Ala'a AL-Zetoun,}
\author[a]{Igor Ostrovskiy,}
\author[a, b]{Peiwen Mi,}
\author[c]{Daniele Perri}

\affiliation[a]{Institute of High Energy Physics, Chinese Academy of Sciences, 19B Yuquan Road, Beijing 100049, China}
\affiliation[b]{School of Physical Sciences, University of Chinese Academy of Sciences, 19A Yuquan Road, Beijing 100049, China}
\affiliation[c]{Institute of Theoretical Physics, Faculty of Physics, University of Warsaw, ul. Pasteura 5, PL-02-093 Warsaw, Poland\\}

\abstract{
Origin and composition of ultra-high energy cosmic rays are still uncertain, particularly the rare events exceeding the Greisen-Zatsepin-Kuzmin cutoff and whose apparent arrival directions point to the Local Void. This work is an in-depth investigation of the possibility that such cosmic rays contain low-mass magnetic monopoles predicted in several recent theoretical models. Using a custom extension to CRPropa, a state-of-the-art code for modeling cosmic ray propagation, monopoles are tracked in the galactic environment under realistic assumptions on mass, magnetic charge, and initial phase-space distributions. 
Our simulations demonstrate that relic MMs follow filamentary ``Galactic magnetic funnels'', producing anisotropic arrival directions at Earth. The expected arrival directions are concentrated in a small region of the sky, which has significant implications for the design and interpretation of future searches. A comparison of the simulated arrival directions with the Pierre Auger Observatory dataset shows that the null hypothesis -- no monopole contribution -- is statistically preferred for the majority of cases. Using a profile-likelihood analysis that incorporates both directional and energy information, we set 90\% C.L. upper limits on the integral MM flux of $\less 2.1\times10^{-23}$ $\mathrm{cm^{-2}s^{-1}sr^{-1}}$. 

\medskip
\textit{Corresponding authors:} 
Igor Ostrovskiy (igor.ostrovskiy@ihep.ac.cn) and 
Daniele Perri (daniele.perri@fuw.edu.pl).
}

\begin{document}
\maketitle

\flushbottom

\section{Introduction}

Ultra-high energy cosmic rays (UHECRs) are astroparticles with energies exceeding 1 EeV~\cite{uhecr_review_2025}.
While tens of thousands of UHECRs have been detected to date, their composition, origin, and acceleration mechanisms are still debated~\cite{batista_review_2019}. 
The observed distribution of arrival directions and the energetic constraints on Galactic acceleration mechanisms suggest the extragalactic origin of the UHECRs~\cite{auger_anisotropy_4EeV_2018, auger_anisotropy_8EeV_2017, hillas_1984}. However, some UHECRs have been observed with energies above the Greisen–Zatsepin–Kuzmin (GZK) cutoff~\cite{gzk_1,gzk_2}, which favors shorter travel distances. While the exact cutoff value depends on the mass composition of the UHECRs -- 30-60 EeV for protons but several thousands of EeV for heavy nuclei -- the latter are subject to other interactions with the CMB, like photodisintegration, that may limit their travel distances~\cite{heavy_nucl_disint_1969}. In either case, the mass composition of UHECRs is currently not well constrained~\cite{mass_composition_report_2017}. A particularly striking case is the recent detection of the Amaterasu particle~\cite{Amaterasu_2023}. Not only its reconstructed energy estimate -- 240 EeV -- far exceeds the GZK cutoff for a proton, but its arrival direction points to the Local Void, an empty region of space between the Local Group of galaxies and nearby large-scale structure filaments~\cite{local_void_2008}. The latter conclusion stands even if the primary particle is an iron nucleus. 

Potential explanations for such occurrences include~\cite{Amaterasu_2023} larger-than-expected foreground magnetic fields and masses of the primaries, or new physics. In particular, it has long been suggested that UHECRs may be magnetic monopoles (MMs)~\cite{crs_are_mms_1996}. MMs are particles with an isolated south or north magnetic pole. Their existence would explain the quantization of the electric charge~\cite{Dirac:1931} and is predicted by Grand Unification Theories (GUT)~\cite{Hooft:74, Polyakov:1974ek}. Incidentally, MMs allow for a natural explanation for the most puzzling features of the highest-energy cosmic rays. The observed strength and coherence lengths of the galactic and intergalactic magnetic fields could accelerate a MM to relativistic velocities as long as its mass is $\lesssim$10$^{8}$ TeV/c$^2$~\cite{perri_intergalactic_acc_2023}. MMs would not be subject to cutoff-like effects due to negligible expected interaction with the CMB~\cite{Cho_2024} and their continued acceleration in the presence of magnetic field offsetting possible energy losses, even on Galactic scales. Moreover,  if produced in cosmological phase transitions, MMs would be distributed throughout the Universe rather than confined to localized astrophysical accelerators. Such events would therefore not be expected to point back to specific sources. 

At the time when the Ref.~\cite{crs_are_mms_1996} was published, MMs with mass $\lesssim 10^{8}$ TeV/c$^2$ were already motivated by a subclass of GUT models like $SO(10)$ in which symmetry breaking proceeds through one or more intermediate steps.\footnote{An example is provided by the following symmetry-breaking pattern of an initial $SO(10)$ GUT group:
\begin{equation}
SO(10) \longrightarrow SU(4) \otimes SU(2) \otimes SU(2) \longrightarrow SU(3)_c \otimes SU(2) \otimes U_Y(1) .
\end{equation}
The first phase transition might occur at GUT scales $10^{15}$ GeV, close to the inflationary scales, breaking the $SO(10)$ symmetry into the cross product group $SU(4) \otimes SU(2) \otimes SU(2)$. Hence, the second transition, where the $SU(4) \otimes SU(2) \otimes SU(2)$ group breaks into the Standard Model group, might occur at much lower scales. One expects the resulting MM mass to be of order $\tilde{g} v$, where $\tilde{g}$ is the coupling of the group in $SU(4) \otimes SU(2) \otimes SU(2)$ that is broken into the hypercharge $U(1)$ group and $v$ the vacuum expectation value of the second symmetry breaking.} However, a subsequent work argued that MMs are unlikely to be highest-energy cosmic ray events based on poor compatibility with observations of the predicted energy spectrum and arrival directions~\cite{crs_are_not_mms_1999}. 

Since then, several developments have taken place. Firstly, more accurate predictions of arrival directions are now possible because the quantitative understanding of the galactic magnetic field has been improved substantially in the past two decades, thanks to numerous Faraday rotation and starlight polarization measurements~\cite{new_gala_fields1, new_gala_fields2, new_gala_fields3}. Secondly, an array of Beyond Standard Model theories has been proposed in recent years that contain finite-energy MM solutions with masses as low as $\sim10^{0}$--10$^2$ TeV/c$^2$ and fundamental magnetic charges up to six times larger than the Dirac charge~\cite{CHO:1997, cho:2015, Ellis:2016, mavromatos:2017, Arunasalam:2017, Mavromatos:2018, Hung:2020, Shafi:2022, bps_cho_mason_2018, Alonso:2025rkk}. Thirdly, a previously unexplored MM production mechanism, the Schwinger process \cite{Schwinger:1951nm,Affleck:1981ag,AFFLECK1982509}, has been gaining substantial attention, suggesting that low-mass MMs could be produced by astroparticle sources of strong magnetic fields in the present epoch~\cite{Gould_PRL_2017}. Lastly, the number of detected UHECRs has increased significantly in the past twenty years, improving the ability to statistically compare the observed UHECRs with different hypotheses. In particular, the Pierre Auger Observatory (PAO) conducted a dedicated search assuming the MMs interact with atmosphere only through electromagnetic interactions~\cite{PierreAuger:2016imq}. Under this assumption, a relativistic MM traversing the atmosphere would produce a characteristic continuous signal in the PAO detectors, distinct from the signal of a hadronic shower by its intensity and geometry. The absence of such events allowed PAO to set 90\% C.L. upper limits on the flux of ultra-relativistic MMs, reaching $\sim 10^{-20}$ cm$^{-2}$s$^{-1}$sr$^{-1}$ for Lorentz factors $\gamma > 10^8$ \cite{PierreAuger:2016imq}. 

MMs interacting only electromagnetically represent only a small subclass of all the possible MM models. Examples of this subclass are point-like Dirac MMs \cite{Dirac:1931} and 't Hooft-Polyakov MMs with charges 3 or 6 times the fundamental magnetic charge \cite{Alonso:2025rkk}.
In contrast, most theoretical constructions predict particles with non-trivial internal QCD structure. For such MMs, interactions with ordinary matter are expected to include QCD processes, especially if the MM carries a colored core or couples to gluons. These QCD interactions could cause a composite MM to initiate a hadronic shower in the atmosphere that closely resembles that of a heavy nucleus~\cite{WICK2003663}, potentially evading the event selection criteria designed to isolate electromagnetic MM signatures. Consequently, the stringent Auger limits may still leave the question of whether some of the UHECR events with puzzling features could be MMs.

In light of the above developments, this work revisits the possibility that MMs could contribute to the UHECR flux by simulating low-mass ($\lesssim 100~\mathrm{TeV/c^2}$) MMs in the galactic environment and their expected distribution of arrival directions and kinetic energies at Earth.
In Section~\ref{sec:methods} the simulation framework and assumed initial conditions are described. Section~\ref{sec:sim_results} presents the simulation results. Section~\ref{sec:anisotropy} discusses the implication of the anisotropic arrival directions for cosmic ray detectors. In Section~\ref{sec:Auger_comparison}, the simulation results are compared to the PAO's UHECR dataset to obtain model-independent constraints on the MM flux.
The paper then concludes by discussing the limitations of the study and its implications for the current and future MM searches.

\section{Methods}
\label{sec:methods}

This section describes the simulation framework and inputs used for the simulation -- models of the Galactic magnetic field and initial MM distributions considered in two scenarios: relic MMs accelerated in galactic magnetic fields in the early stages of their evolution, and relic MMs accelerated in intergalactic magnetic fields.

\subsection{Magnetic monopole simulation framework}

The simulation of MMs was conducted using CRPropa~3.2.1, a publicly available Monte Carlo simulation framework developed for studying the propagation of cosmic rays and secondary particles in galactic and extragalactic environments~\cite{crpropa_2022}. 
CRPropa is designed with a modular and extensible architecture, allowing users to customize simulation pipelines and incorporate additional physics modules. Although MMs are not implemented in the official release, the open-source nature of the code makes it possible to introduce new particle types and corresponding interaction models. In this work, dedicated extensions were developed to describe the propagation dynamics of MMs within astrophysical magnetic fields. In particular, we implemented three custom modules for the simulation of MMs. 

The first two modules are the modified versions of the CRPropa built-in propagation algorithms, originally designed to handle electrically charged particles propagating in magnetic fields. One of the modules solves the equations of motion using the Boris push method~\cite{Boris1970}, and the other one uses the Cash-Karp method~\cite{CashKarp1990}. These modules have been extended to work for particles with both electric and magnetic charges, or dyons. The propagation of a particle in a magnetic field is implemented by solving the modified Lorentz force equation (in SI units):
\begin{equation}
    \frac{d\mathbf{p}}{dt} \equiv \mathbf{F} = g\mathbf{B} + q\mathbf{v}\times\mathbf{B},
\end{equation}
where $\mathbf{p}$ is the particle relativistic momentum, $g$ is the magnetic charge of the particle in Ampere-meter units, $\mathbf{B}$ is the magnetic field at the position of the particle, $q$ is the electric charge of the particle, and $\mathbf{v}$ is its velocity. The second term is only relevant for dyons, which are not considered in this work. The base propagation modules in CRPropa assume the particles are ultra-relativistic, i.e., $\beta \approx 1$, whereas the two custom propagation modules allow the particles to have velocities less than the speed of light. Both the base and custom modules allow for adaptive size of the path-length increment (step).

The third module simulates the radiative losses of an accelerating magnetically charged particle. The relativistic generalization of the Larmor formula and electromagnetic duality transformation give the radiated power for a MM as (in SI units)
\begin{equation}
    P=\frac{\mu_0g^2}{6\pi m^2c^3}\gamma^2
    \left(
        \left(
            \frac{d\mathbf{p}}{dt}
        \right)^2
        -\frac{1}{c^2}
        \left(
            \frac{dE}{dt}
        \right)^2
    \right)
    =\frac{\mu_0g^2}{6\pi m^2c^3}
    \left(
        \mathbf{F}^2+\frac{\gamma^2}{c^2}|\mathbf{v}\times\mathbf{F}|^2
    \right),
\end{equation}
where $P$ is the radiated power, $\mu_0$ is the permeability of free space, $m$ is the particle mass, $\gamma$ is the Lorentz factor, and $E$ is the particle energy. The energy of the particle is modified at the end of each step, with the radiated energy subtracted off determined by the second expression of the above formula times the step size.

The modules were validated by propagating MMs in a uniform magnetic field and comparing them to the analytical solutions of the equations of motion. For a relativistic MM propagating in a uniform magnetic field, the trajectory is the same as a relativistic electrically charged particle propagating in a uniform electric field, which follows a hyperbolic cosine curve. Full details of the custom modules and their detailed validations will be subject to a separate publication.

\subsection{Simulation assumptions and settings}
\label{sec:settings}

MMs could have been produced in the early universe during the symmetry breaking stage via the Kibble mechanism~\cite{Kibble_1976,Rajantie_2002}. Given the low mass of MMs considered in this work, they would have been produced after the inflationary epoch and not expected to overclose the universe, with the expected density depending on the vacuum expectation value of the theory. Recently, it was also suggested that MMs could have been produced via the Schwinger effect in primordial magnetic fields~\cite{Kobayashi_2021,Kobayashi_2023}. 
Regardless of the production mechanism, the low-mass relic MMs considered in this work would be accelerated by the galactic and intergalactic magnetic fields to relativistic speeds and therefore would not be gravitationally bound to galaxies. Consequently, the MMs are injected uniformly distributed on a 20 kpc sphere centered on the Galaxy with a Lambertian distribution of initial directions.\footnote{A Lambertian distribution corresponds to particle emission from a surface with angular probability density $P(\theta) \propto \cos\theta$, where $\theta$ is the angle between the particle direction and the local surface normal.}

For each simulation, an initial population of $2 \times 10^{10}$ MMs is generated. In the simulation, Earth is positioned 8.2 kpc from the galactic center and 25 pc above the galactic plane, consistent with recent measurements~\cite{earth_distance_from_galactic_center, Blaauw:1960_earth_position_galactic_plane, Siegert:2019_earth_position_galactic_plane}. MMs are propagated in the Galactic field until one of the three conditions is met: a) their distance from Earth is less than 35 pc; b) their distance from the Galactic center exceeds 25 kpc; c) their total travel distance exceeds 50 kpc. In case of the first condition, the MMs are assumed to have reached the Earth and their arrival direction and energy are recorded. For the last two conditions, the MMs are assumed to exit the galaxy. The 35 pc sphere around Earth is chosen to conservatively account for the uncertainty in the Earth's distance from the Galactic center~\cite{earth_distance_from_galactic_center}.

\begin{figure}[htpb]
    \centering
    \begin{subfigure}[b]{\textwidth}
        \centering
        \includegraphics[width=\textwidth]{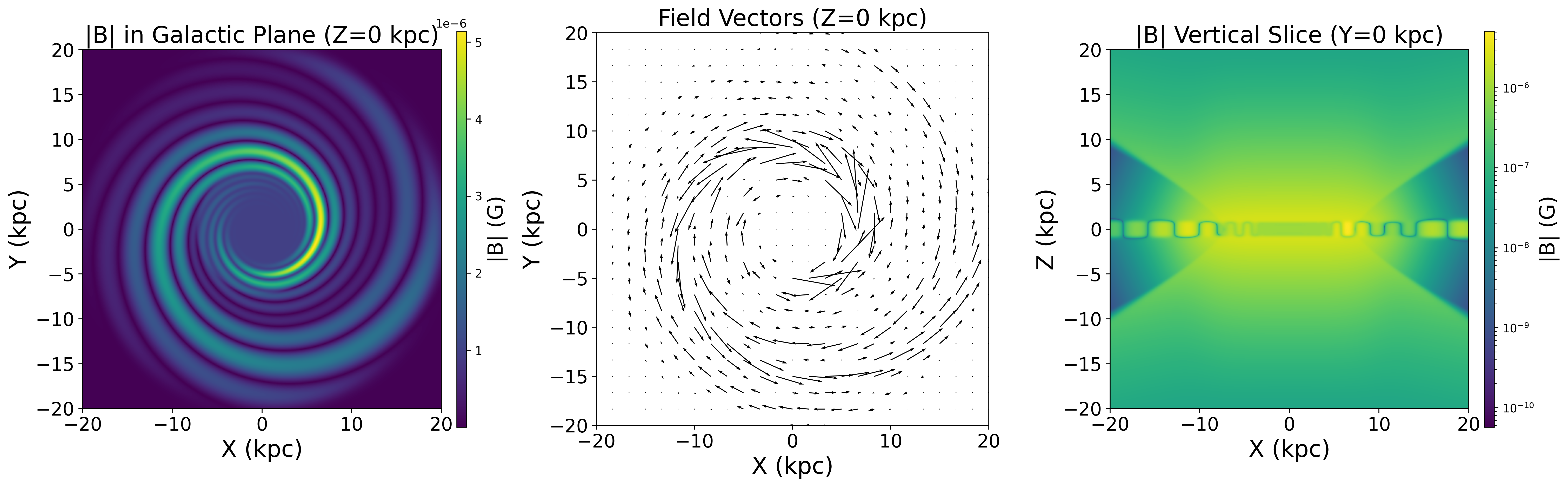}
        \caption{UF23 base model.}
        \label{fig:UF23_base}
    \end{subfigure}
    
    \vspace{0.5em}
    
    \begin{subfigure}[b]{\textwidth}
        \centering
        \includegraphics[width=\textwidth]{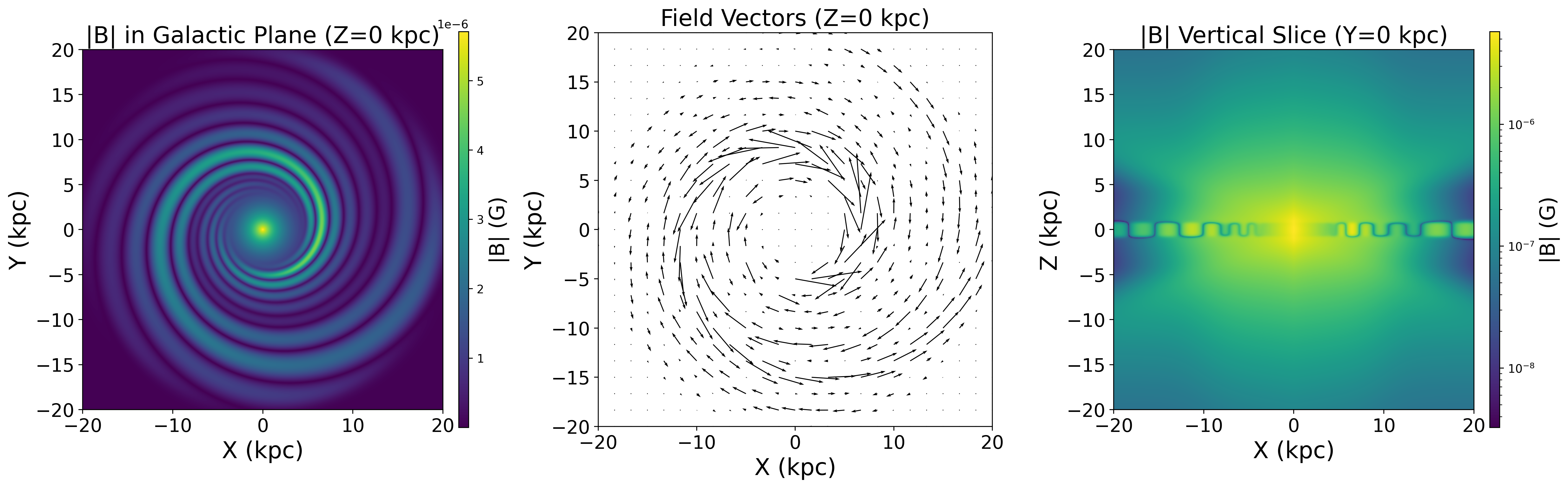}
        \caption{UF23 expX model.}
        \label{fig:UF23_expX}
    \end{subfigure}

    \caption{Comparison of the Galactic magnetic field models adopted for the simulation results shown in this work. The upper panel shows the UF23 base model, which consists of a spiral disk field, an explicit toroidal halo component, and a poloidal X-field. 
    The lower panel shows the UF23 expX variant, which modifies the poloidal component by adopting an exponential dependence for the midplane vertical field instead of the logistic radial cutoff used in the base model.}
    \label{fig:UF23_models}
\end{figure}

The Cash-Karp propagation module is chosen with an adaptive step-size between 0.1 pc and 100 pc, and a tolerance of 10$^{-4}$. MMs are propagated using different models of the galactic magnetic field from the UF23 suite~\cite{UF23_2024}. The suite is an ensemble of eight models that reflect uncertainties due to limitations of the currently available data. 
While the simulations have been run for all eight models, in the following sections we only show the results for the so-called ``base'' and ``expX'' UF23 models as being the most and least similar to the JF12 model~\cite{new_gala_fields1} commonly used earlier. The UF23 base model is the sum of a spiral disk field, an explicit toroidal halo, and a coasting poloidal X-field. The expX variant uses an exponential dependence of the midplane vertical poloidal field instead of the default logistic radial cutoff. Figure~\ref{fig:UF23_models} shows the two magnetic field models. We do not include any turbulent component of the Galactic magnetic field in our simulations, as no updated model of the turbulent components of UF23 is currently available in the literature. However, we discuss the effects of turbulence using the older JF12 model in Appendix~\ref{app:turbulent}.

MMs with masses of $1~\mathrm{TeV}/c^2$ and $100~\mathrm{TeV}/c^2$ are simulated, as representative of the low-mass range. Two magnetic charge values, $g=g_\mathrm{D}$ and $g=6g_\mathrm{D}$ are used, where $g_{\rm D}$ is a Dirac charge, expressed in SI units as $2 \pi \hbar / \left(e \mu_0 c \right)$. This choice represents the two extreme cases among the typical models considered in the literature~\cite{Alonso:2025rkk}. Two initial momentum distributions are used, corresponding to two potential MM acceleration mechanisms, as described in the following section.

\subsection{Initial kinetic energy of relic monopoles}

The initial kinetic energy distribution of MMs is determined by the details of acceleration in cosmic magnetic fields. Here we consider two models of the acceleration, depending on whether the MMs cluster with galaxies at the time of galaxy formation or not. 
The first analyzed case corresponds to MMs at rest in the CMB rest frame at the time of galaxy formation (Scenario I). These MMs will be captured in the gravitational well of the collapsing halo, remaining clustered until the formation of galactic magnetic fields and then accelerated out of their host galaxies. In this case, the MMs will reach the Milky Way with kinetic energies corresponding to the velocities with which they exited the host galaxies. These kinetic energies are estimated by simulations, as described in Section~\ref{sec:GMF_i}. 
In the second scenario, we assume the presence of intergalactic magnetic fields filling the intergalactic voids (Scenario II). In this scenario, the MMs are initially accelerated by magnetic fields in intergalactic voids and therefore never cluster with galaxies. These MMs are assumed to have a monochromatic energy distribution and isotropic directions when they enter the Milky Way, as discussed in Section~\ref{sec:intergalactic} below.

\subsubsection{Scenario I: acceleration in galactic fields}
\label{sec:GMF_i}

In the scenario of acceleration by galactic magnetic fields, we assume that the MMs entering the Milky Way were previously accelerated in the galaxies in which they originally clustered. To estimate their initial kinetic-energy distribution upon entering the Milky Way, we further assume that the magnetic history of their host galaxies was broadly similar to that of the Milky Way.

The origin of magnetic fields in spiral galaxies is commonly attributed to dynamo mechanisms that amplify primordial seed fields \cite{Haverkorn2015,Brandenburg_2023}, operating since the epoch of structure formation. However, the detailed evolution of these processes remains an active area of research, as it depends sensitively on the properties of the initial seed field.
If the initial seed field is sufficiently strong, a large-scale dynamo alone can account for the magnetic field strengths observed in present-day galaxies \cite{Martin_Alvarez_2021}. In this case, the large-scale dynamo both amplifies and organizes the magnetic field over a characteristic timescale of $\sim 1~\mathrm{Gyr}$.
Conversely, if the seed field is weak, the large-scale dynamo is not sufficient by itself to generate the observed field amplitudes. In this case, a much faster small-scale dynamo, operating on timescales of $\sim 10^7~\mathrm{yr}$, drives the initial amplification~\cite{Beck_2012}. At the end of this phase, the magnetic field remains highly turbulent, with a coherence length of only $\sim 10~\mathrm{pc}$. A subsequent large-scale dynamo phase is therefore still required to build up the coherent, ordered structures observed in galaxies today. After the dynamo process reaches saturation, the magnetic field strength is generally assumed to remain approximately constant. As the case of strong seed field is usually linked to the existence of primordial intergalactic fields (which is already included in Scenario II), we do not consider it here. Therefore, in Scenario I, it is assumed that the MMs are initially accelerated by the turbulent fields produced by the small-scale dynamo amplification. We then set the conditions for the initial kinetic energy of the MMs by accelerating them in the turbulent-only component of the JF12 Galactic magnetic field module and record the energies with which they exit the galaxy. 

The initial position distribution of MMs in the host galaxy depends on their masses and charges, as well as on the dynamical timescale of halo formation~\cite{Dunsky:2018mqs}. For sufficiently large charges and small masses, MMs are expected to trace the baryonic component during halo collapse and ultimately share its spatial distribution. In contrast, for smaller charges and larger masses, interactions with baryons are too weak for MMs to thermalize with the baryonic matter, and they instead retain the primordial distribution associated with the dark matter halo. However, identifying which one of the two options is the most apt to describe each simulation depends on the details of galaxy evolution and will not be discussed here.
Therefore, this work considers both possibilities separately, adopting the results of Ref.~\cite{2020MNRAS.494.4291C} for a baryon-like distribution, and a standard Navarro–Frenk–White (NFW) profile~\cite{Navarro:1995iw,Navarro:1996gj} for a dark matter-like distribution.

\begin{figure}[htpb]
  \centering
  \begin{subfigure}[b]{0.49\textwidth}
    \centering
    \includegraphics[width=\textwidth]{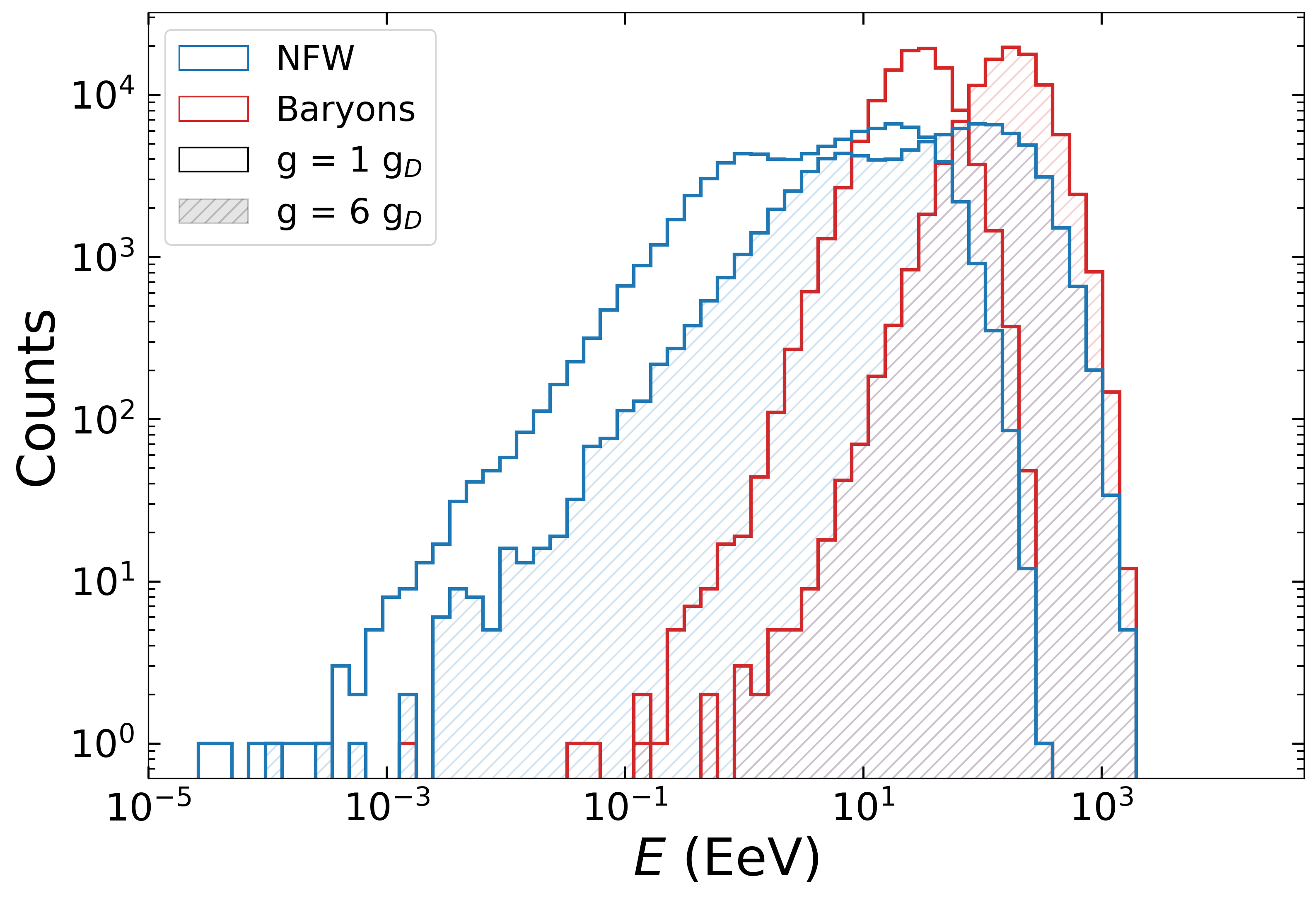}
        
  \end{subfigure}
  \hfill
  \begin{subfigure}[b]{0.49\textwidth}
    \centering
    \includegraphics[width=\textwidth]{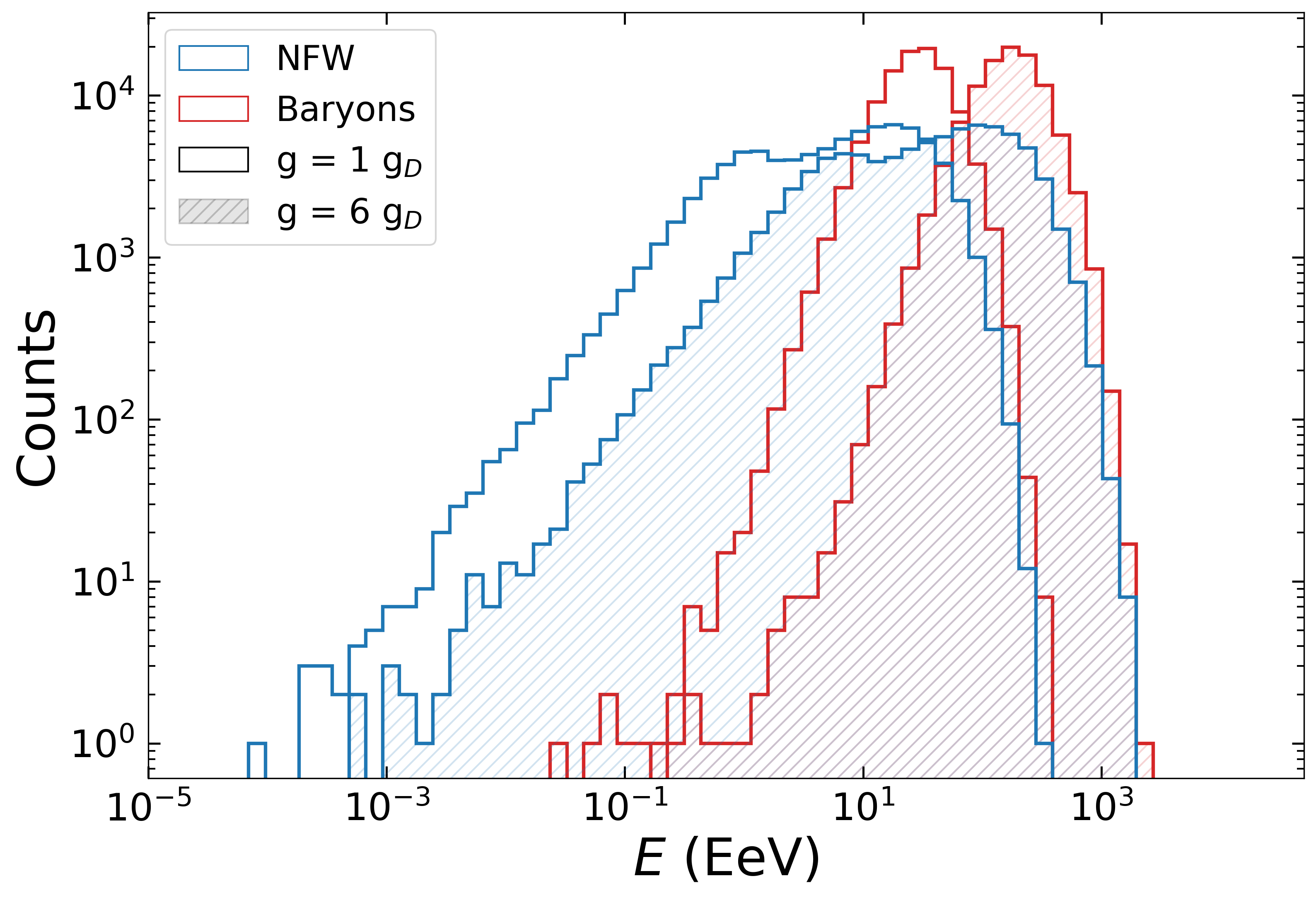}
    
  \end{subfigure}
   \caption{Kinetic energy distributions of MMs upon exiting their host galaxy, assumed here to be analogous to the Milky Way. The turbulent component of the JF12 magnetic field model is used as a representative for the earlier epoch. MMs with a mass of 1 TeV/c$^2$ (left plot) and 100 TeV/c$^2$ (right plot) are used in the simulation. MMs with initial positions following the NFW and baryon profiles are shown in blue and red, respectively. Histograms with hatched lines correspond to a magnetic charge of 6 $g_{\mathrm{D}}$, whereas unhatched histograms represent singly charged MMs.}
    \label{fig:gmf_init}
\end{figure}

Figure~\ref{fig:gmf_init} shows the resulting initial kinetic-energy distributions for the cases of a $1~\mathrm{TeV}/c^2$ (left plot) and $100~\mathrm{TeV}/c^2$ (right plot) MM masses.
The distributions for the two different mass choices are practically identical. In the figure, the red curves correspond to simulations assuming a baryonic MM initial position distribution, while the blue curves correspond to an NFW one. Unhatched histograms represent singly charged MMs ($g = g_{\mathrm{D}}$), whereas hatched histograms correspond to the case $g = 6 g_{\mathrm{D}}$. As expected, increasing the magnetic charge shifts the kinetic-energy distribution toward higher energies. In particular, for the baryonic distribution, the kinetic-energy distribution exhibits a peak around $35~\mathrm{EeV}$ for $g = g_{\mathrm{D}}$ and $176~\mathrm{EeV}$ for $g = 6g_{\mathrm{D}}$. On the other hand, for the NFW distribution, the corresponding modes are around $13~\mathrm{EeV}$ ($18~\mathrm{EeV}$) for a MM mass of $100~\mathrm{TeV/c^2}$ ($1~\mathrm{TeV/c^2}$) with $g = g_{\mathrm{D}}$, and $92~\mathrm{EeV}$ for $g = 6g_{\mathrm{D}}$ in both mass scenarios. In all cases, the majority of MMs have arrival energies within approximately one order of magnitude of the peak value.
On the other hand, assuming an NFW profile leads to a significantly broader kinetic-energy distribution. These differences arise naturally from the different spatial distributions. MMs following the baryonic distribution are preferentially located in the central galactic regions, where magnetic fields are stronger, and therefore experience more efficient acceleration. In contrast, MMs distributed according to an NFW profile extend farther into the galactic halo, where magnetic fields are substantially weaker, resulting in lower typical acceleration and a broader range of final kinetic energies.

The distributions obtained in this section are subsequently used in Section~\ref{sec:sim_results} to sample the initial kinetic energies of MMs entering the Milky Way in Scenario I.

\subsubsection{Scenario II: acceleration in intergalactic fields}
\label{sec:intergalactic}

In the case of acceleration by intergalactic magnetic fields, a simple estimate of the accumulated energy can be obtained by multiplying  the MM charge with the average magnetic field strength and the coherence length of the field~\cite{perri_intergalactic_acc_2023}. Using recent estimates for the intergalactic field strength, this yields kinetic energies as large as 10$^4$ EeV~\cite{intergalactic_fields1, intergalactic_fields2}. 
Following Ref.~\cite{perri_intergalactic_acc_2023}, the intergalactic magnetic fields are modeled with coherence length~$\lambda_{\mathrm{I}}$, by dividing the universe into cells of uniform field, with each cell having a size~$\lambda_{\mathrm{I}}$. Each cell is then associated with the same value for the field amplitude~$B_{\mathrm{I}}$ and the velocity of the MMs after a Hubble time is computed. Assuming that the MMs are initially at rest at $t = 0$ and always accelerated to relativistic velocities, we use the expression from Ref.~\cite{perri_intergalactic_acc_2023} to estimate the MMs kinetic energy in the Milky Way rest frame after being accelerated for a Hubble time:
\begin{equation}
     E_{\rm kin} = m (\gamma - 1) \sim m \gamma \sim \mathrm{min.} \left\{  gB_{\mathrm{I}} \left(\frac{\min\left(\lambda_{\mathrm{I}},1/H_0\right)}{H_0}\right)^{1/2},  \frac{B_{\mathrm{I}}^2}{8 \pi F_{\mathrm{I}}} \right\},
\label{eq:v_igmf}
\end{equation}
where $F_{\mathrm{I}}$ is the intergalactic MM flux per area per time per solid angle, in the CMB rest frame.
In the region of the parameter space that is of interest for UHECR detectors, i.e., $F \gtrsim 10^{-21} \mathrm{cm^{-2}s^{-1}sr^{-1}}$, the expression for the kinetic energy is always given by the second term of Eq.~\eqref{eq:v_igmf} for the intergalactic magnetic field parameters allowed by current constraints \cite{batista_review_2019}. 

Whether or not the acceleration in the intergalactic magnetic field can significantly affect the motion of the MMs depends on whether the kinetic energy in Eq.~\eqref{eq:v_igmf} is larger or smaller than the kinetic energy that MMs at rest in the CMB frame have in the Milky Way frame. The case of a significant intergalactic acceleration is verified for masses
\begin{equation}
\label{eq:mass_l}
    m \lesssim \frac{B_I^2}{4 \pi v_p^2 F_I} \sim 10^7~\mathrm{GeV}~\left( \frac{B_I}{10^{-14}~\mathrm{G}}\right)^2 \left( \frac{F_I}{10^{-18}~\mathrm{cm^{-2}s^{-1}sr^{-1}}}\right)^{-1},
\end{equation} 
where $v_p \sim 10^{-3}$ is the peculiar velocity of the Milky Way in the CMB rest frame. If the MM mass is larger than the expression in Eq.~\eqref{eq:mass_l}, the acceleration in intergalactic magnetic fields is not sufficient to make the MMs unclustered with the Milky Way at the time of Galaxy formation. In this case, the initial kinetic energy can be estimated as described in Section~\ref{sec:GMF_i} and is already included in Scenario I. Therefore, for Scenario II, we consider only the combination of the MM mass and flux and amplitude of the intergalactic magnetic fields for which the condition in Eq.~\eqref{eq:mass_l} is verified, which implies that MMs do not cluster. Their expected kinetic energy at the entrance of the Milky Way is then set by Eq.~\eqref{eq:v_igmf} and will have a monochromatic spectrum.

For the simulation, four benchmark values of the magnetic field amplitude and MM flux are adopted, with each scenario corresponding to a specific initial kinetic energy. 
The different choices are listed in the first four entries of Table~\ref{tab:ike_III}.
In all the cases, the kinetic energy gained in intergalactic magnetic fields is expected to be smaller than in Galactic fields. 
For comparison, a fifth scenario was also considered, corresponding to the case of an intergalactic magnetic field with the highest amplitude admitted by current constraints and a MM flux which corresponds to an initial kinetic energy much larger than the expected contribution from Galactic magnetic fields. The details of the parameters adopted in this scenario are shown in the last entry of Table~\ref{tab:ike_III} (Scenario IIe).
\begin{table}[htpb]
    \centering
    \begin{tabular}{|| c | c | c | c ||}
         \hline 
          Scenario & $B_{\rm I}~(\mathrm{G})$ & $F_{\rm I}~(\mathrm{cm^{-2}s^{-1}sr^{-1}})$ & $E_{\rm kin}^i~(\mathrm{GeV})$ \\ [0.5ex] 
         \hline 
         IIa & $10^{-14}$ & $10^{-18}$ & $6$ \\ [0.5ex] 
         \hline  
         IIb & $10^{-14}$ & $10^{-20}$ & $6 \times 10^2$ \\ [0.5ex] 
         \hline 
         IIc & $10^{-10}$ & $10^{-18}$ & $6 \times 10^8$ \\ [0.5ex] 
         \hline  
         IId & $10^{-10}$ & $10^{-20}$ & $6 \times 10^{10}$ \\ [0.5ex] 
         \hline  
         IIe & $10^{-9}$  & $10^{-21}$ & $6 \times 10^{13}$ \\ [0.5ex] 
         \hline
    \end{tabular}
    \caption{Different choices of the intergalactic magnetic field amplitude, MM flux and the relative initial kinetic energy assumed for the MMs in the simulations of Scenario II.}
    \label{tab:ike_III}
\end{table}
\FloatBarrier

\section{Results of the simulations}
\label{sec:sim_results}

The results of the simulations show that relic MMs exhibit anisotropic arrival distributions at Earth despite their initial positions and velocities being sampled from uniform and isotropic (Lambertian) distributions. This behavior arises because their initial kinetic energies are sufficiently low for their propagation to be dominated by the Galactic magnetic field, forcing them to follow its field lines. As long as the initial kinetic energy remains small compared with the energy subsequently gained through magnetic acceleration, the qualitative behavior of the trajectories is unchanged. The main difference between the scenarios is therefore only quantitative and is determined by the typical kinetic energy with which the MMs enter the Galaxy. Lower initial kinetic energies lead to stronger magnetic confinement and consequently more pronounced anisotropies in the arrival directions, whereas higher initial kinetic energies allow larger deviations from the field lines, producing broader arrival-direction distributions. In this section, the results of the different simulations are discussed in detail. 

\begin{figure}[htpb]
  \centering
  \begin{subfigure}[b]{0.49\textwidth}
    \centering
    \includegraphics[width=\textwidth]{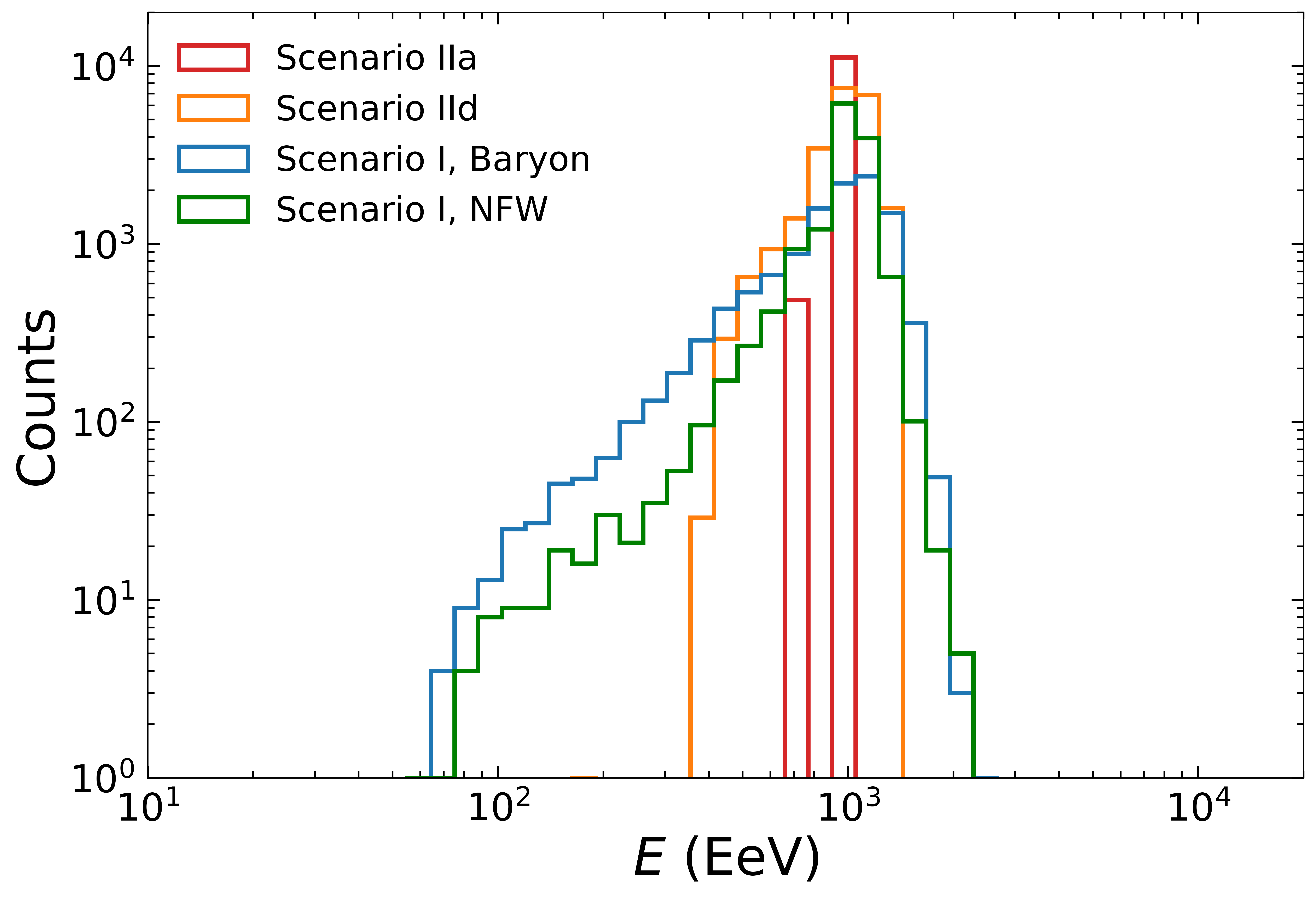}
    \caption{$g=g_{\rm D}$}
    \label{fig:ek_f_1gD}
  \end{subfigure}
  \hfill
  \begin{subfigure}[b]{0.49\textwidth}
    \centering
    \includegraphics[width=\textwidth]{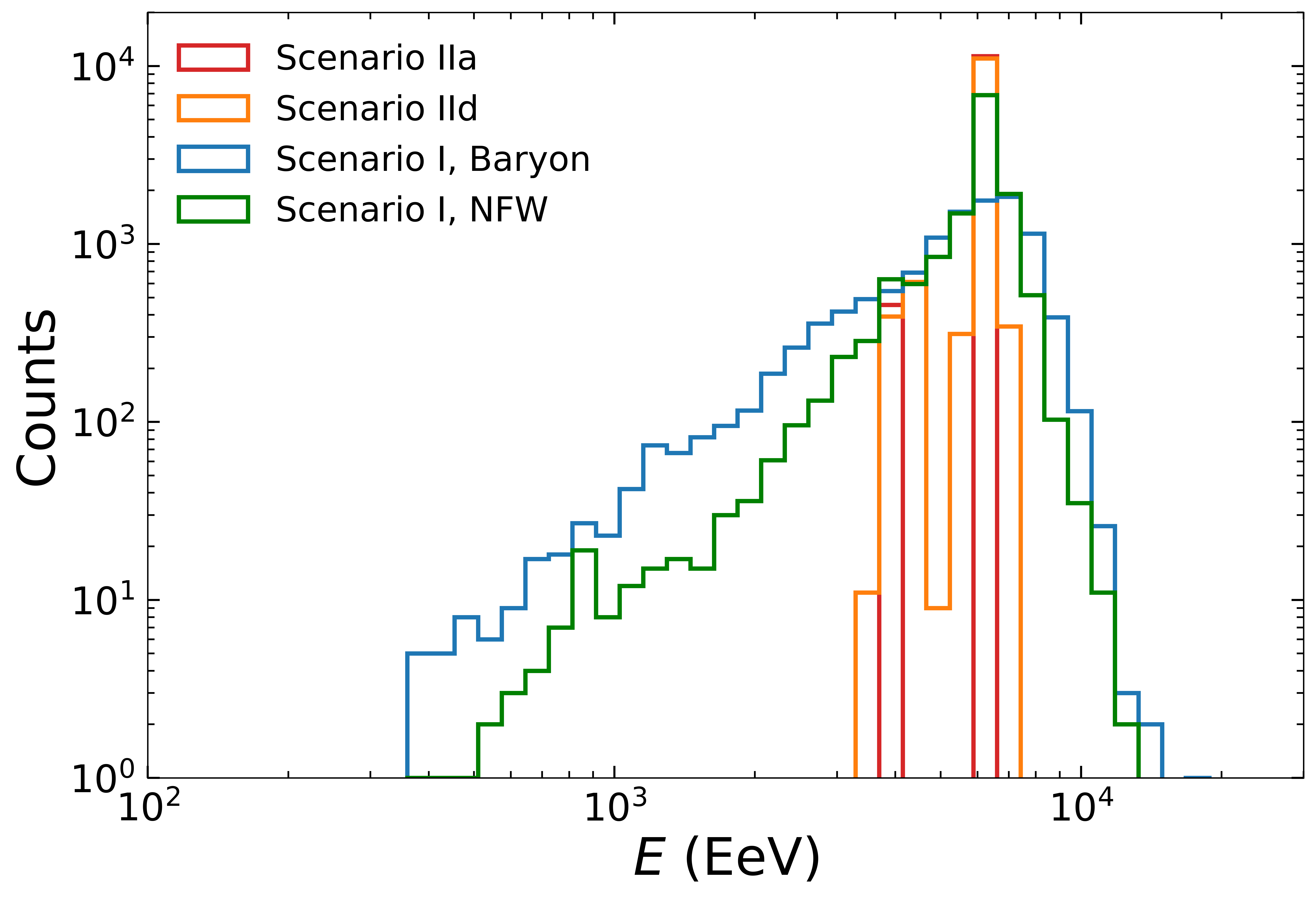}
    \caption{$g=6g_{\rm D}$}    
    \label{fig:ek_f_6gD}
  \end{subfigure}
\caption{Arrival kinetic-energy distributions of MMs at Earth for a mass of $100~\mathrm{TeV}/c^2$, obtained using the UF23 base Galactic magnetic-field model. The left and right panels correspond to $g=g_{\rm D}$ and $g=6g_{\rm D}$, respectively. The curves show the Scenario I baryonic (blue) and NFW (green) initial distributions, together with the Scenario II cases IIa (red) and IId (orange).}
    \label{fig:ek_base}
\end{figure}

Figure~\ref{fig:ek_base} shows the simulated distributions of MM arrival kinetic energies at Earth for MMs with a mass of $100~\mathrm{TeV}/c^2$. 

\begin{figure}[htpb]
  \centering
  \begin{subfigure}[b]{\textwidth}
    \centering
    \includegraphics[width=\textwidth]{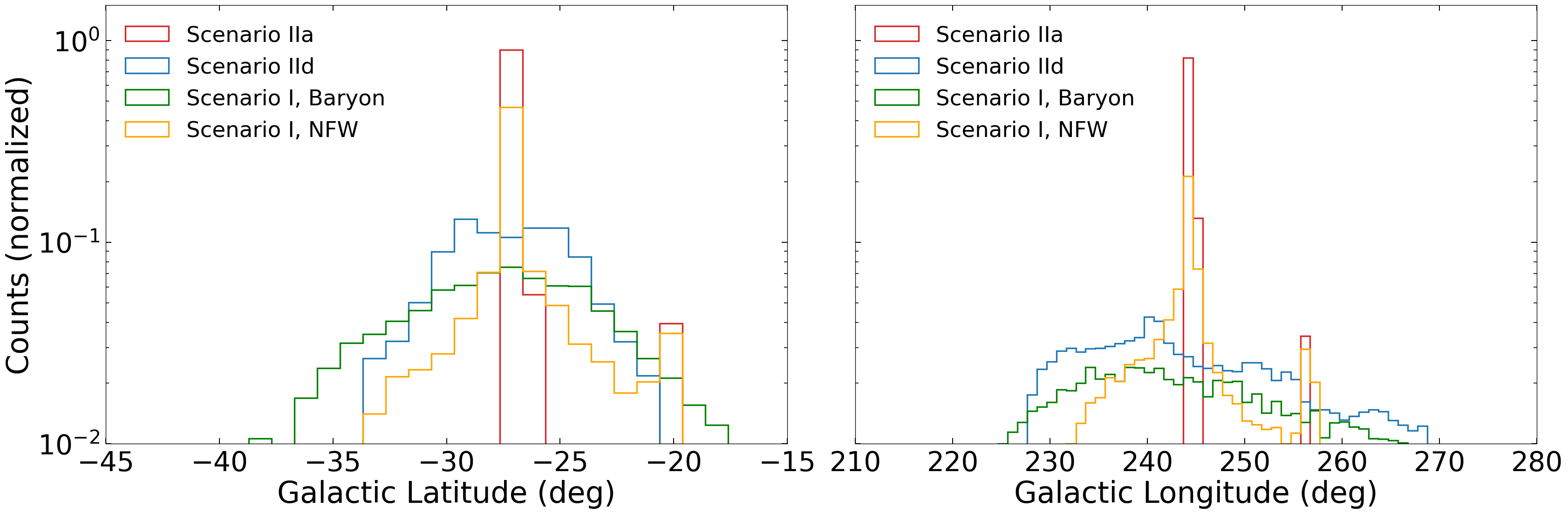}
    \caption{$g=g_{\rm D}$}
    \label{fig:angles_f_1gD}
  \end{subfigure}
  \hfill
  \begin{subfigure}[b]{\textwidth}
    \centering
    \includegraphics[width=\textwidth]{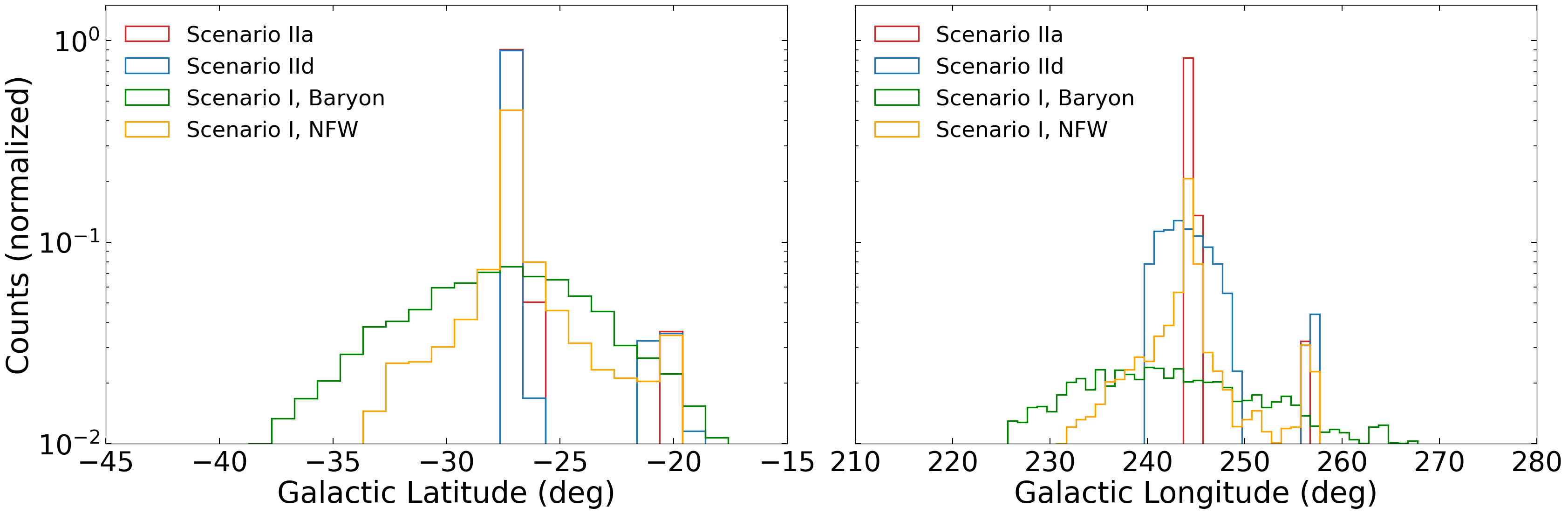}
    \caption{$g=6g_{\rm D}$}
    \label{fig:angles_f_6gD}
  \end{subfigure}

  \caption{Arrival angular distributions of MMs at Earth for a mass of $100~\mathrm{TeV}/c^2$, obtained using the UF23 base Galactic  magnetic-field model. The top and bottom panels correspond to  $g=g_{\rm D}$ and $g=6g_{\rm D}$, respectively. The curves show the  Scenario I baryonic (blue) and NFW (green) initial distributions, together with the Scenario II cases IIa (red) and IId (orange).}
  \label{fig:angles_base}
\end{figure}

The distributions are shown for the initial kinetic energy of Scenario I (MMs initially clustered with baryons, blue curve, or following the initial NFW halo profile, green curve) and for two representative cases of Scenario II (case IIa, red curve, and case IId, orange curve). The left panel corresponds to $g=g_{\rm D}$, while the right panel shows the case $g=6g_{\rm D}$. All simulations were performed using the UF23 base model of the Galactic magnetic field. The corresponding angular distributions of MMs at Earth are shown in Figure~\ref{fig:angles_base}.


The simulations show that the position of the distribution peak is determined entirely by the magnetic charge. In particular, the preferred arrival kinetic energy is found to be approximately $1.1\times10^{3}~\mathrm{EeV}$ for MMs and anti-MMs with $|g|=g_{\rm D}$ and $6.5\times10^{3}~\mathrm{EeV}$ for $|g|=6g_{\rm D}$.
These values are around one order of magnitude larger than simplified analytical estimates presented in~\cite{Kobayashi_2023}, where the average kinetic energy was approximated as $E\sim g B_{G} \sqrt{R_G \lambda_G} \sim (g/g_{\rm D}) \times 200 ~\mathrm{EeV}$, with $B_{G} = 2 \times 10^{-6}~\mathrm{G}$, $R_G = 10~\mathrm{kpc}$, and $\lambda_G = 1~\mathrm{kpc}$. 
Although the choice of the initial distribution has little influence on the location of the energy peak, it noticeably affects the width of the distribution. In Scenario I, the baryonic profile produces a broader energy spectrum than the NFW profile because the typically higher initial velocities allow the MMs to explore a wider range of trajectories through the Galactic magnetic field. A similar trend is observed in Scenario II. For the lowest initial kinetic energy (Scenario IIa), the arrival energies are concentrated within a very narrow range around the preferred value, whereas the larger initial kinetic energies of Scenario IId lead to a broader distribution, reflecting a larger spread in the magnetic energy gained during Galactic propagation. Overall, these results demonstrate that the characteristic arrival energy is largely insensitive to the initial kinetic energy, provided that the latter remains subdominant to the energy acquired from the Galactic magnetic field.

\begin{figure}[htpb]
  \centering
  \begin{subfigure}[b]{\textwidth}
    \centering
    \includegraphics[width=\textwidth]{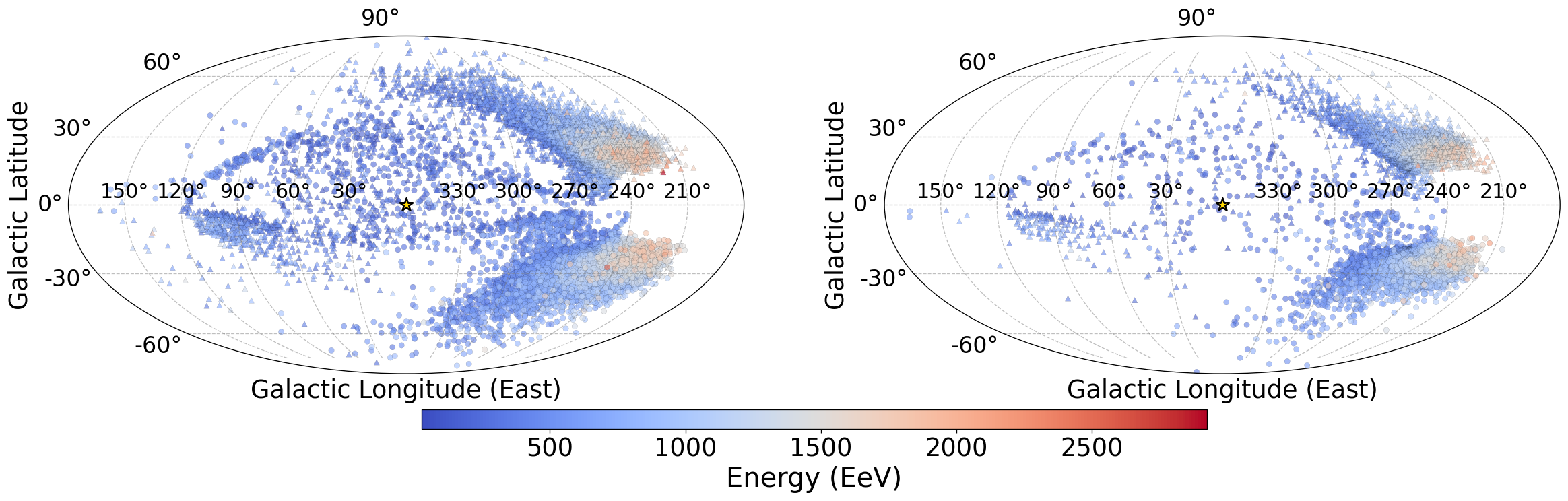}
    \caption{ a) Baryon, $g=g_{\rm D}. $\hspace{5cm}   b) NFW, $g=g_{\rm D}$.}
    \label{fig:II_1gD}
  \end{subfigure}
  \vspace{0.5cm}
  \begin{subfigure}[b]{\textwidth}
    \centering
    \includegraphics[width=\textwidth]{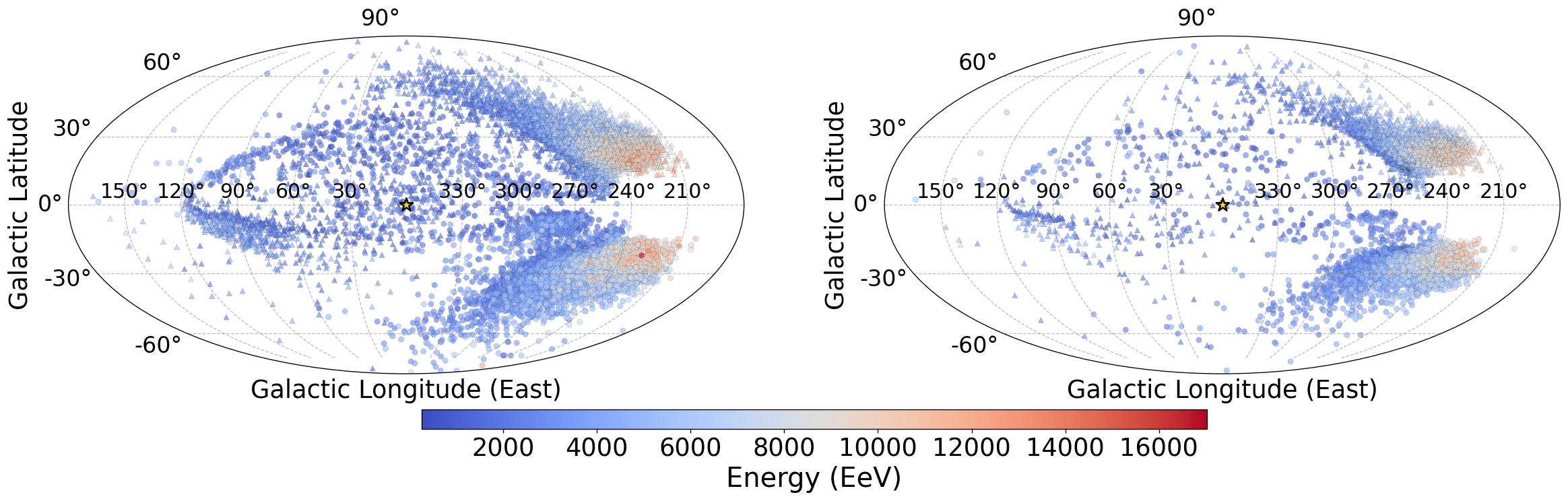}
    \caption{ c) Baryon, $g=6g_{\rm D}$. \hspace{5cm}  d) NFW, $g=6g_{\rm D}$.}
    \label{fig:II_6gD}
  \end{subfigure}
\caption{Skymap of the arrival directions of relic MMs (circles) and anti-MMs (triangles) originating from populations clustered with baryons (left) and following the NFW profile (right) before being accelerated away from their host galaxies. Results are shown for a MM mass of $100~\mathrm{TeV}/c^2$ and two different choices of the magnetic charge, $g=g_{\rm D}$ and $g=6g_{\rm D}$, propagated through the UF23 base Galactic magnetic-field model. The yellow star marks the Galactic Center (GC).}
    \label{fig:scII_skymaps}
\end{figure}

\begin{figure}[htpb]
  \centering
  \begin{subfigure}[b]{\textwidth}
    \centering
    \includegraphics[width=\textwidth]{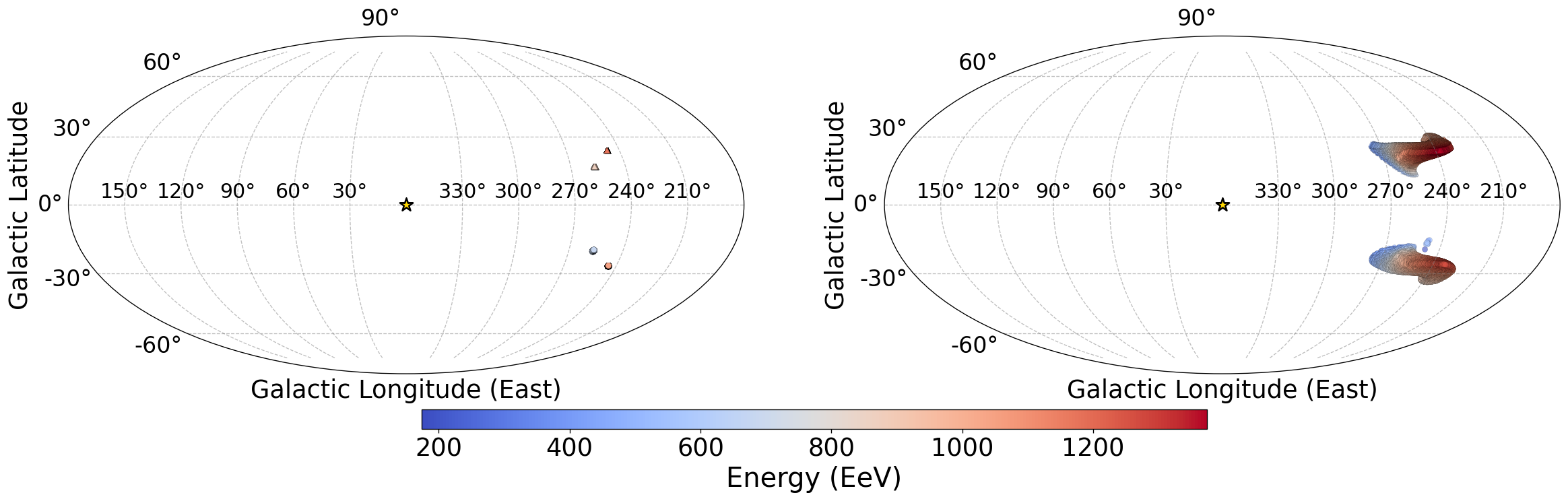}
    \caption{ a) Scenario IIa, $g=g_{\rm D}$\hspace{3cm}
              b) Scenario IId, $g=g_{\rm D}$}
    \label{fig:III_1gD}
  \end{subfigure}
  \vspace{0.5cm}
  \begin{subfigure}[b]{\textwidth}
    \centering
    \includegraphics[width=\textwidth]{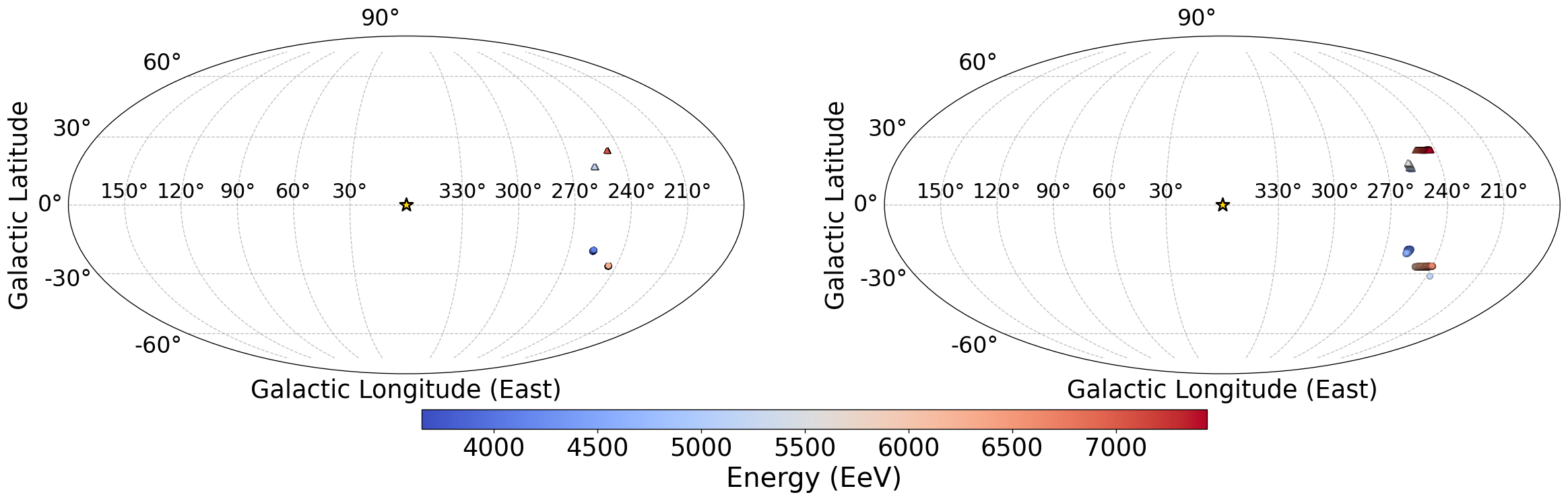}
    \caption{ c) Scenario IIa, $g=6g_{\rm D}$. \hspace{3cm}
              d) Scenario IId, $g=6g_{\rm D}$.}
    \label{fig:III_6gD}
  \end{subfigure}
  \caption{Skymap of the arrival directions of relic MMs (circles) and anti-MMs (triangles) reaching Earth from an initially isotropic distribution, considering selected initial kinetic energies in Scenario II and two different choices of the magnetic charge, $g=g_{\rm D}$ and $g=6g_{\rm D}$. The simulations were performed for a MM mass of $100~\mathrm{TeV}/c^2$ using the UF23 base Galactic magnetic-field model. The yellow star marks the Galactic Center (GC). }
  \label{fig:skymap_III}
\end{figure}

The simulated arrival-direction skymap distributions for both the initial distributions of Scenario I (baryonic and NFW) are shown in Figure~\ref{fig:scII_skymaps}. 
\captionsetup[subfigure]{labelformat=empty}
The corresponding arrival-direction skymap distributions for Scenario II are shown in Figure~\ref{fig:skymap_III} for the two benchmark initial kinetic energies considered in this section, Scenarios IIa and IId.
\captionsetup[subfigure]{labelformat=empty}
The figures present the results for MMs with 100 TeV/c$^2$ mass propagated through the UF23 base model for two magnetic charges ($g=g_{\rm D}$ and $g=6g_{\rm D}$). Both MMs (circles) and anti-MMs (triangles) are shown, as they are always expected to be produced in the same amount. 

A characteristic feature of both scenarios is the clear separation between the arrival directions of MMs and anti-MMs. Because particles with opposite magnetic charge propagate along the Galactic magnetic field in opposite directions, they reach Earth from opposite regions of the sky. The degree of anisotropy, however, depends strongly on the initial kinetic energy. In Scenario I, the relatively large initial kinetic energies of a fraction of the simulated MMs allow noticeable deviations from individual magnetic field lines, so that the Galactic magnetic field imprints a pronounced but relatively broad large-scale anisotropy on the arrival directions. Consistent with the arrival-energy distributions, the baryonic initial profile produces a broader angular distribution than the NFW profile. In contrast, the lower initial kinetic energies of Scenario II keep the MMs more tightly confined to the Galactic magnetic field lines, resulting in a stronger anisotropy. In both the scenarios, the arrival directions are concentrated within narrow regions of the sky, with MMs arriving predominantly from below the Galactic plane (Galactic latitude $\sim-30^\circ$ and longitude $\sim250^\circ$) and anti-MMs from the approximately mirror-symmetric region above the plane.

The observed anisotropy, together with the very narrow distributions of the arrival kinetic energy, is a direct consequence of the relic MMs having sufficiently low initial kinetic energies to become magnetically confined by the Galactic magnetic field. Rather than propagating along nearly rectilinear trajectories, these MMs are forced to follow the magnetic field lines, which effectively act as ``Galactic magnetic funnel''. The consequence of this behavior is discussed in Section~\ref{sec:anisotropy}.
The separation of MMs and anti-MMs into opposite hemispheres follows naturally from the opposite signs of their magnetic charges and their propagation along the coherent Galactic magnetic field in opposite directions.
Finally, increasing the magnetic charge further reduces the angular spread of the arrival directions, consistent with the increasingly dominant role of the Galactic magnetic field in determining the MM dynamics.

The simulations were also performed for Scenarios IIb and IIc. For all considered values of the magnetic charge, no significant differences with respect to Scenario IIa were found. It was also checked with dedicated simulations that reducing the MM mass to 1 TeV/c$^2$ leaves the resulting distributions essentially unchanged. 
Therefore, throughout the remainder of the paper, we refer to the results as generically valid for low-mass monopoles ($m \lesssim 100~\mathrm{TeV}/c^2$) without explicitly specifying the mass considered for each simulation result.
Finally, the impact of adopting different Galactic magnetic field model was investigated. In particular, the corresponding simulations for the UF23 expX model are presented in Appendix~\ref{app:expX}. The results exhibit only minor differences compared with those obtained using the UF23 base model, indicating that the conclusions are qualitatively independent of the specific Galactic magnetic field model adopted.

As previously mentioned, as an additional check a simulation was repeated with the initial kinetic energy increased to $6 \times 10^{4}~\mathrm{EeV}$, a value larger than the potential energy of the Galactic field and which corresponds to the extreme values of the intergalactic magnetic field amplitude allowed by observations. As expected, in this case the distribution of arrival directions at Earth becomes nearly isotropic over the sky, in clear contrast with the lower-energy cases discussed above (see Figure~\ref{fig:IG1_extreme} in Appendix~\ref{app:IIIf}). This behavior is expected because the kinetic energy carried by the MMs before entering the Galaxy is now much larger than the energy gained from acceleration in the Galactic magnetic field. Consequently, their trajectories are no longer significantly guided by the Galactic magnetic field lines. Instead, the MMs propagate almost rectilinearly through the Galaxy, preserving their original directions of motion.

\section{Implications of an anisotropic flux for experimental limits}
\label{sec:anisotropy}

\begin{figure}[htpb]
  \centering
  \begin{subfigure}[b]{0.45\textwidth}
    \centering
    \includegraphics[
        width=\textwidth,
        trim=60 100 60 10,
        clip
    ]{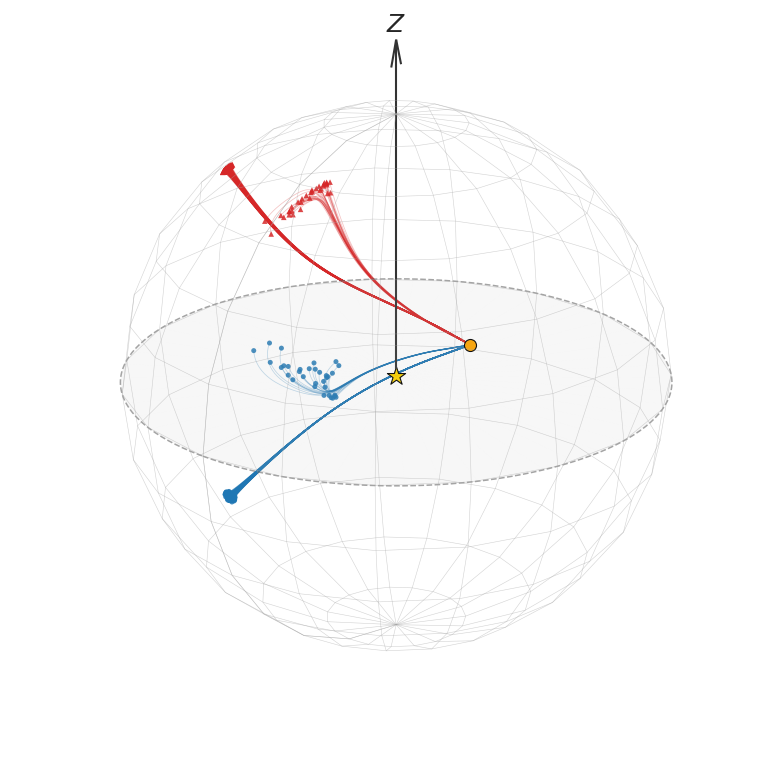}
    \caption{(a) Scenario IIc}
    \label{fig:funnel_c}
  \end{subfigure}
  \hfill
  \begin{subfigure}[b]{0.45\textwidth}
    \centering
    \includegraphics[
        width=\textwidth,
        trim=60 100 60 10,
        clip
    ]{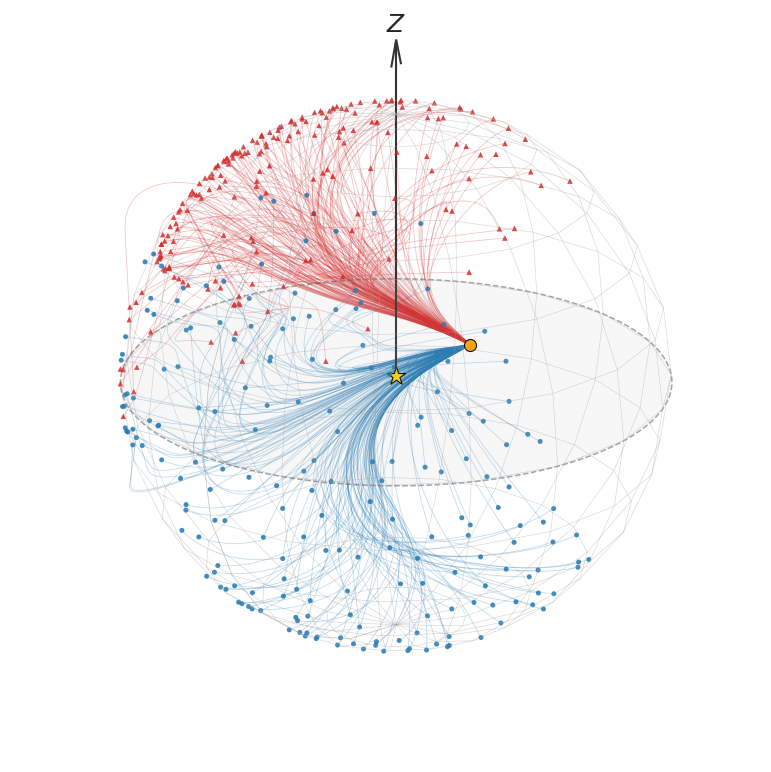}
    \caption{(b) Scenario IId}    
    \label{fig:funnel_d}
  \end{subfigure}
\caption{3D trajectory plot of selected MMs (blue) and anti-MMs (red) with mass 100 TeV/c$^2$ and $|g|=g_{\rm D}$, simulated in Scenario IIc (left) and Scenario IId (right) with UF23 base field model. Only trajectories reaching Earth are shown. Earth and the Galactic center are indicated by orange circle and yellow star, respectively. The corresponding source positions on the 20 kpc injection sphere are marked by blue circles and red triangles for MMs and anti-MMs, respectively. The clustering of the source positions and convergence of the trajectories toward Earth illustrate the "Galactic magnetic funnel" effect induced by the Galactic magnetic field.}
    \label{fig:funnel}
\end{figure}

An important question to ask about the observed anisotropy of the MMs' arrival directions is whether MMs experience any focusing, whereby particles from a wide range of initial locations are funneled by the Galactic field onto a common arrival region. Figure~\ref{fig:funnel} shows the trajectories of selected MMs simulated in Scenario IIc and IId, demonstrating that a focusing effect is indeed present. 
We call this effect the ``Galactic magnetic funnel'' and discuss its implications for cosmic-ray experiments in this section.

To quantify the effect of the Galactic magnetic funnel, we compute the probability $P_{\rm Earth}$ for an MM to reach the Earth in the given scenario and define the funneling ratio $\eta$ as the ratio of this probability to the corresponding probability $P_{\rm iso}$ obtained by propagating MMs in the absence of a Galactic magnetic field.\footnote{The subscript indicates that, in the absence of a Galactic magnetic field, the MM distribution at Earth is isotropic.}
For an isotropic distribution of MMs on a surrounding sphere, only MMs arriving from a limited fraction of the total solid angle can ultimately reach the Earth. In the presence of magnetic fields, however, this accessible fraction can be modified by magnetic deflection and focusing. Consequently, a flux limit derived under the assumption of an isotropic MM flux, $F<F_*$, as is the case for current constraints from terrestrial experiments, must be corrected by dividing by the funneling ratio $\eta$ to account for the anisotropic propagation induced by the Galactic magnetic field. Therefore, the corrected expression for the flux bounds is\footnote{An additional complication for the results of Scenario II is that the energy depends on the flux too. This has to be taken into account when recasting the constraints for different limits on the MM flux at the Earth.}
\begin{equation}
\label{eq:flux_corrected}
    F < F_* \eta^{-1} .
\end{equation}

\begin{figure}[htpb]
    \centering
    \includegraphics[width=0.75\textwidth]{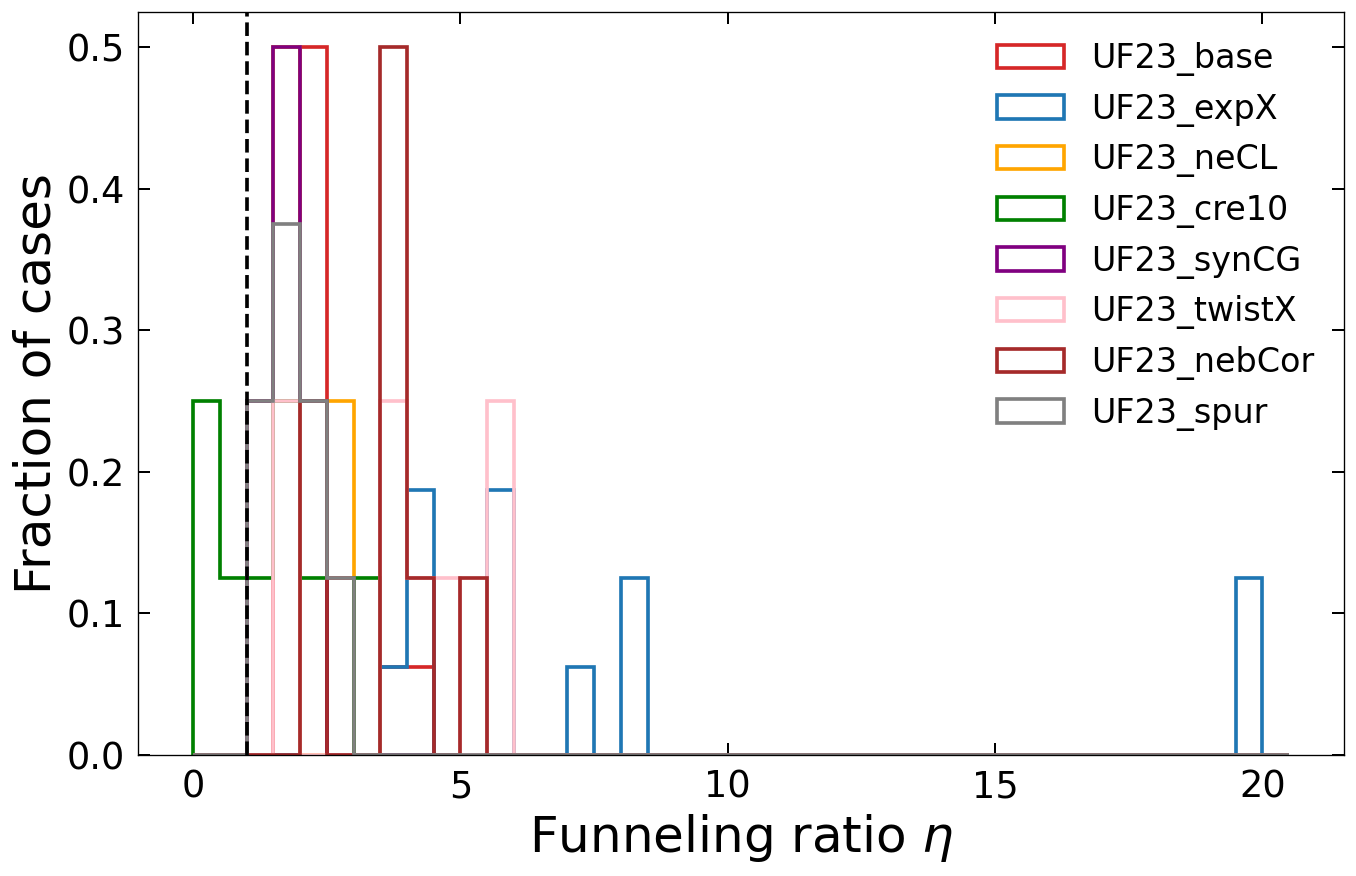}
    \caption{Funneling ratio, $\eta$, for all simulations performed in this work. The colors identify the field models, while the multiple entries associated with each model correspond to the combinations of initial kinetic energy, an MM mass of 1 TeV/$c^2$, and magnetic charge of $g_D$ (every model) and $6 g_D$ (only base and expX), including both MMs and anti-MMs. The vertical dashed line marks $\eta$ = 1, corresponding to no focusing effect. Values above (below) correspond to an enhancement (suppression) of the number of MMs reaching the Earth relative to the no-field case.}
    \label{fig:funnel_factor}
\end{figure}

The funneling ratios for the simulation results shown in Section~\ref{sec:sim_results} are summarized in Table~\ref{tab:funnel_ratio}, while their distribution for the eight UF23 field models is shown in Figure~\ref{fig:funnel_factor}. 
\begin{table}[htpb]
    \centering
    \small
    \setlength{\tabcolsep}{3pt}
    \renewcommand{\arraystretch}{1.05}
    
    \begin{subtable}[t]{0.49\textwidth}
    \centering
    \begin{tabularx}{\linewidth}{
        ||C{1.10}|C{1.25}|C{0.65}|C{0.95}|C{1.05}||
    }
    \hline
    $E_{\rm k, i}$ & Charge ($g_{\rm D}$) & Field & $N_{\rm hit}$ & $\eta$ \\ [0.5ex]
    \hline
    \hline
    \multirow{8}{*}{Baryons}
        & \multirow{2}{*}{1}
        & base & 11549 & 1.88(3) \\ [0.5ex]
        \cline{3-5}
        &
        & expX & 12606 & 2.06(3) \\ [0.5ex]
        \cline{2-5}
        & \multirow{2}{*}{-1}
        & base & 12534 & 2.04(3) \\ [0.5ex]
        \cline{3-5}
        &
        & expX & 13124 & 2.14(3) \\ [0.5ex]
        \cline{2-5}
        & \multirow{2}{*}{6}
        & base & 11424 & 1.86(3) \\ [0.5ex]
        \cline{3-5}
        &
        & expX & 12496 & 2.04(3) \\ [0.5ex]
        \cline{2-5}
        & \multirow{2}{*}{-6}
        & base & 12438 & 2.03(3) \\ [0.5ex]
        \cline{3-5}
        &
        & expX & 13195 & 2.15(3) \\ [0.5ex]
    \hline
    \multirow{8}{*}{NFW}
        & \multirow{2}{*}{1}
        & base & 14190 & 2.31(4) \\ [0.5ex]
        \cline{3-5}
        &
        & expX & 26361 & 4.30(6) \\ [0.5ex]
        \cline{2-5}
        & \multirow{2}{*}{-1}
        & base & 15631 & 2.55(4) \\ [0.5ex]
        \cline{3-5}
        &
        & expX & 49354 & 8.05(11) \\ [0.5ex]
        \cline{2-5}
        & \multirow{2}{*}{6}
        & base & 13998 & 2.28(4)  \\ [0.5ex]
        \cline{3-5}
        &
        & expX & 26193 & 4.27(6)  \\ [0.5ex]
        \cline{2-5}
        & \multirow{2}{*}{-6}
        & base & 15506 & 2.53(4)  \\ [0.5ex]
        \cline{3-5}
        &
        & expX & 49071 & 8.00(11)  \\ [0.5ex]
    \hline
    \end{tabularx}
    \caption{Scenario I}
    \label{tab:funnel_II}
    \end{subtable}
    \hfill
    \begin{subtable}[t]{0.49\textwidth}
    \centering
    \begin{tabularx}{\linewidth}{
        ||C{1.10}|C{1.25}|C{0.65}|C{0.95}|C{1.05}||
    }
    \hline 
    $E_{\rm k, i}$ & Charge ($g_{\rm D}$) & Field & $N_{\rm hit}$ & $\eta$ \\ [0.5ex] 
    \hline 
    \hline 
    \multirow{8}{*}{IIa}
        & \multirow{2}{*}{1}
        & base & 11671 & 1.90(3)  \\ [0.5ex]
        \cline{3-5}
        & & expX & 33941 & 5.53(8)  \\ [0.5ex]
    \cline{2-5}
        & \multirow{2}{*}{-1}
        & base & 12520 & 2.04(3)  \\ [0.5ex]
        \cline{3-5}
        & & expX & 120634 & 19.67(26) \\ [0.5ex]
    \cline{2-5}
        & \multirow{2}{*}{6}
        & base & 11827 & 1.93(3)  \\ [0.5ex]
        \cline{3-5}
        & & expX & 34310 & 5.59(8)  \\ [0.5ex]
    \cline{2-5}
        & \multirow{2}{*}{-6}
        & base & 12703 & 2.07(3)  \\ [0.5ex]
        \cline{3-5}
        & & expX & 120880 & 19.70(26) \\ [0.5ex]
    \hline
    \multirow{8}{*}{IId}
        & \multirow{2}{*}{1}
        & base & 22749 & 3.71(5) \\ [0.5ex]
        \cline{3-5}
        & & expX & 23525 & 3.84(6) \\ [0.5ex]
    \cline{2-5}
        & \multirow{2}{*}{-1}
        & base & 26558 & 4.33(6) \\ [0.5ex]
        \cline{3-5}
        & & expX & 25704 & 4.19(6) \\ [0.5ex]
    \cline{2-5}
        & \multirow{2}{*}{6}
        & base & 12726 & 2.07(3) \\ [0.5ex]
        \cline{3-5}
        & & expX & 34968 & 5.70(8)  \\ [0.5ex]
    \cline{2-5}
        & \multirow{2}{*}{-6}
        & base & 13745 & 2.24(3) \\ [0.5ex]
        \cline{3-5}
        & & expX & 44669 & 7.28(10) \\ [0.5ex]
    \hline
    \end{tabularx}
    \caption{Scenario II}
    \label{tab:funnel_III}
    \end{subtable}
    \caption{Number of $100~\mathrm{TeV/c^2}$ MMs that reach the Earth and the funneling ratios for different choices of the initial kinetic energy, magnetic field model, and magnetic charge.}
    \label{tab:funnel_ratio}
\end{table}
Although the majority of the cases across all eight models lie within $1 \lesssim \eta \lesssim 6$, there are a few outliers. In particular, the UF23 expX model produces the broadest distribution and includes the extreme case with $\eta \sim 20$, whereas the cre10 model yields $\eta < 1$, corresponding to magnetic de-focusing rather than focusing. Nevertheless, the majority of simulations support the existence of magnetic funnels with a positive focusing effect, which significantly enhances the constraining power of terrestrial experiments.

\begin{figure}[htpb]
    \centering
    \includegraphics[width=\textwidth]{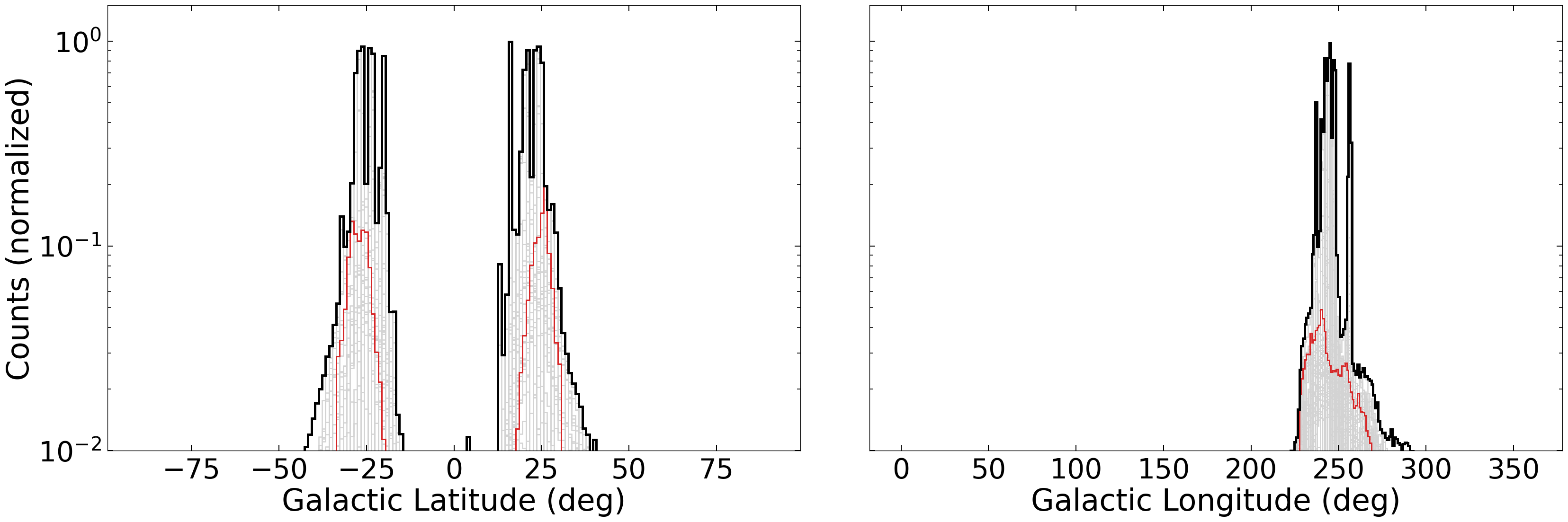}
    \caption{Arrival angular distributions of MMs and anti-MMs propagated through all eight field models of the UF23 suite. Simulations with an MM mass of 1 TeV/c$^2$ and $|g| = g_D$ (every model) and $6 g_D$ (only base and expX) are included. Each light grey curve represents a distinct simulation configuration, while the black envelope marks the combined distribution. The red curve highlights the UF23 base model for scenario IId with  $|g| = g_D$. The concentration of events within relatively narrow angular intervals demonstrates that the predicted arrival directions are strongly anisotropic, with the anisotropy being robust across the full range of Galactic magnetic field model variations.}
    \label{fig:all_angular}
\end{figure}

A second important consequence of the MM anisotropic distribution is that experiments that have an acceptance substantially larger than the effective arrival region predicted by the simulations do not necessarily provide stronger constraints on the MM flux than detectors with a smaller acceptance that nonetheless fully cover the expected arrival region. As shown in Figure~\ref{fig:all_angular}, the arrival directions of MMs and anti-MMs propagated through all the UF23 models are concentrated in a relatively small region of Galactic longitude (bottom panel) and latitude (top panel) angles.
The envelope (black line) -- which still remains restricted to a relatively small region of the sky -- is indicative of the current level of uncertainty in the modeling of the Galactic magnetic field. 
The figure shows that the results are not qualitatively affected by different choices for the Galactic magnetic field model.

\begin{figure}[htpb]
    \centering
    \begin{subfigure}[b]{0.49\textwidth}
        \centering
        \includegraphics[width=\linewidth]{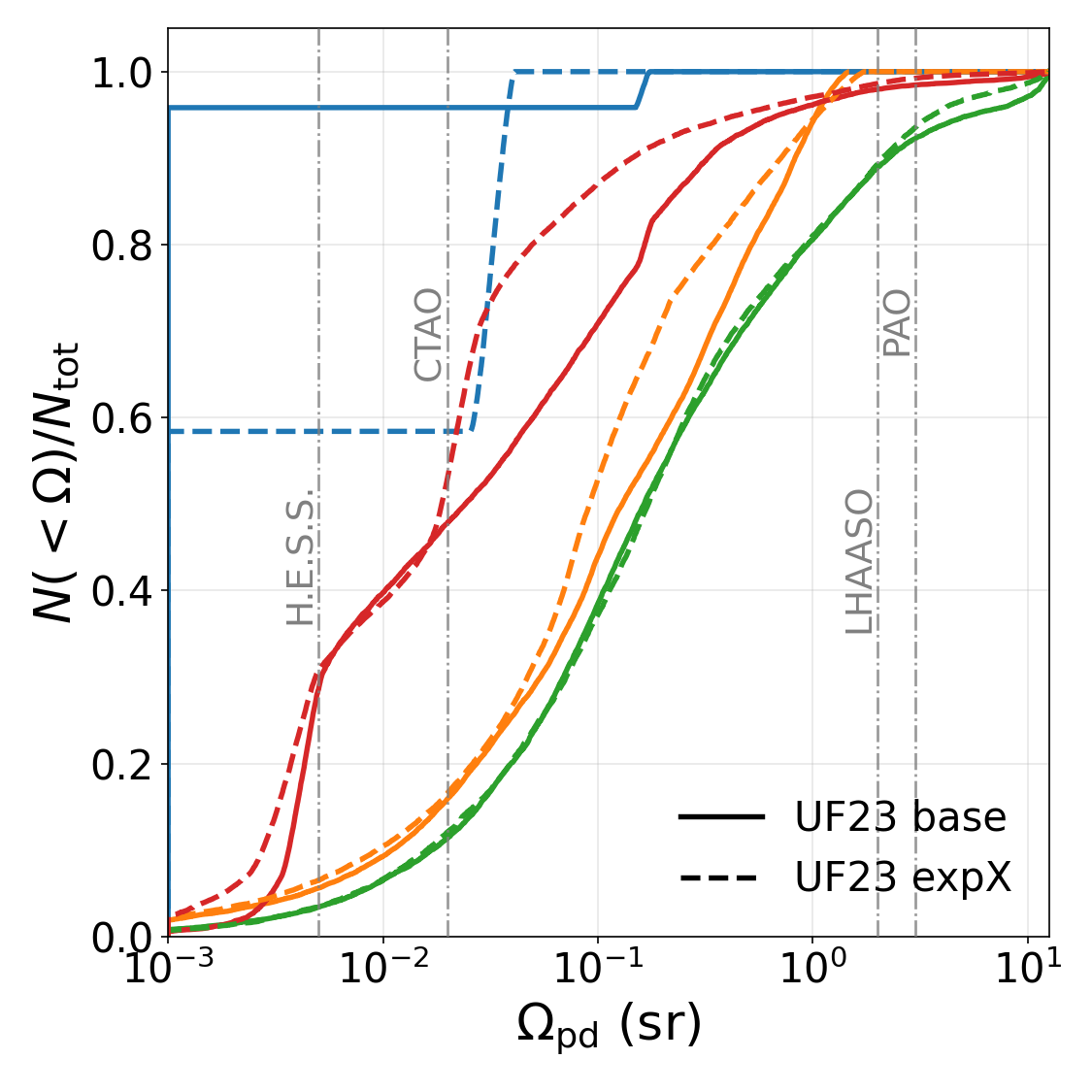}
        \caption{(a) $g = g_{\rm D}$}
        \label{fig:solid_angle_1gD}
    \end{subfigure}
    \hfill
    \begin{subfigure}[b]{0.49\textwidth}
        \centering
        \includegraphics[width=\linewidth]{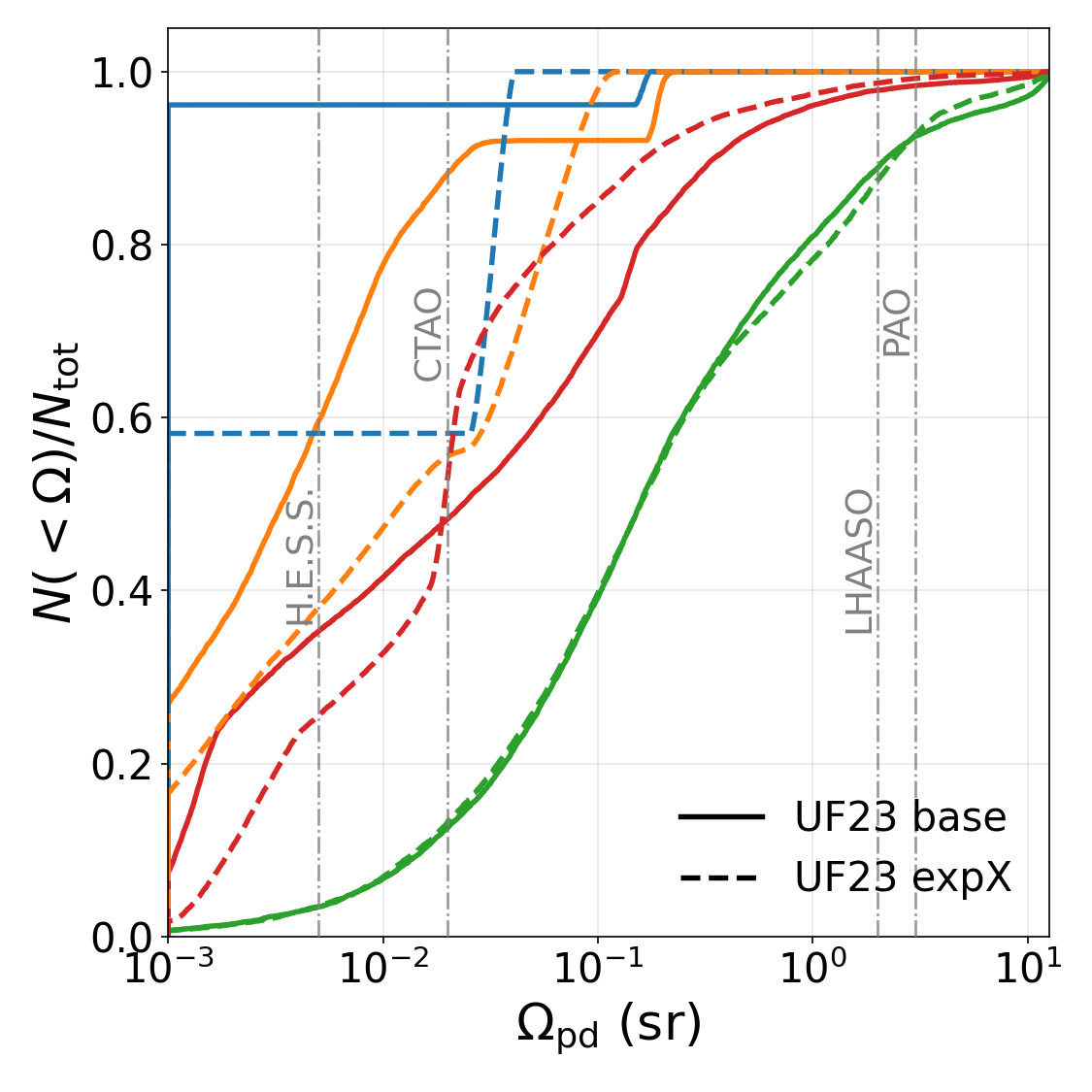}
     \caption{(b) $g = 6g_{\rm D}$}
     \label{fig:solid_angle_6gD}
    \end{subfigure}
    \caption{Fraction of low-mass MMs reaching Earth as a function of the solid angle around the preferred arrival direction, $\Omega_{\rm pd}$, shown for different initial kinetic energies, Galactic magnetic-field models, and two values of the MM charge. Only MMs with north magnetic charge are shown in the figure. Solid (dashed) curves correspond to the UF23 base (expX) field model, while different colors correspond to different initial distributions (green: I-baryons, red: I-NFW, blue: IIa, orange: IId). Dotdashed gray lines correspond to the approximate field of view of current or future cosmic ray observatories (H.E.S.S. \cite{HESS:2018pbp}, CTAO \cite{CTAConsortium:2017dvg}, LHAASO \cite{DiSciascio:2016rgi}, PAO \cite{2015172}).}
    \label{fig:solid_angle}
\end{figure}

On the other hand, Figure~\ref{fig:solid_angle} shows the fraction of events in the simulations as a function of the solid angle around the Galactic coordinates corresponding to the preferred arrival directions at Earth, as discussed in Section~\ref{sec:sim_results}. 
The figure shows low-mass MMs for two different choices of the magnetic charge: $g=g_{\rm D}$ (left) and $g=6g_{\rm D}$ (right). For each initial kinetic-energy scenario (shown in different colors), results with the UF23 base (solid line) and expX (dashed line) models are shown. 
For readability purposes, the anti-MM distribution is not included in the figure. The results shown in the figure confirm that, depending on the initial kinetic energy model, the expected arrival direction distribution at Earth can be highly concentrated. Moreover, the comparison between the two different values of the MM charge confirms the results of Section~\ref{sec:sim_results}, where it was observed that the distribution of the arrival directions for the two charge choices is significantly different only in the case of Scenario IId. The approximate fields of view of current and future cosmic ray observatories (H.E.S.S. \cite{HESS:2018pbp}, CTAO \cite{CTAConsortium:2017dvg}, LHAASO \cite{DiSciascio:2016rgi}, PAO \cite{2015172}) are also shown in the picture in dotdashed gray vertical lines.

The results of this section have important implications for future MM searches in cosmic rays. In particular, detectors with smaller acceptance but higher precision, or whose performance relies less on model-dependent assumptions, may ultimately achieve more stringent constraints than larger instruments with lower precision or greater model dependence. Such a conclusion follows from the fact that the field of view required for a detector to capture a significant fraction of the total Galactic MM flux is substantially smaller than that required for an isotropic arrival direction distribution. This effect is highlighted in Figure~\ref{fig:solid_angle}, where the fraction of the predicted MM signal captured by experiments such as H.E.S.S. and CTAO is not significantly smaller than that captured by experiments with fields of view orders of magnitude larger, such as LHAASO and PAO.

The above results neglect the effects of magnetic field turbulence, as this field component is not implemented in the UF23 suite of Galactic field models currently available in CRPropa. As shown in Appendix~\ref{app:turbulent} for the case of an older field model -- JF12 -- turbulence may broaden the arrival-direction distribution of the MMs in scenarios with low initial kinetic energy. 
A rough estimate of the effect can be obtained by comparing the width of the distributions in Figure~\ref{fig:JF12_angular} with and without the turbulent component. The excess widths due to turbulence ($\sim$4-5 and $\sim$6-8 deg for latitude and longitude, respectively) could then be added in quadrature to the width of the envelop angular distributions in Figure~\ref{fig:all_angular}. In both cases, the envelop distributions are only broadened by a couple of degrees, still retaining the characteristically anisotropic shape. Once suitable updated models of the Galactic turbulent magnetic field become available, their effect could be rigorously included in the results, but the above estimate suggests the effect is relatively mild.

\section{Shower-model-independent flux bounds from Auger UHECR dataset}
\label{sec:Auger_comparison}

In this section, we derive bounds on the MM flux independent of the modeling of the atmospheric shower produced by MMs by comparing their expected arrival-direction distribution and energy with the observed UHECR data from the PAO detector.

\subsection{The Auger dataset}

PAO is the largest cosmic-ray detector in the Southern Hemisphere.
It combines a large Surface Detector (SD) array of water-Cherenkov tanks with Fluorescence Detectors (FD)~\cite{Abraham2004}. The SD continuously records extensive air showers, while the FD provides calorimetric measurements used to calibrate the  SD energy scale. Together, these complementary detectors enable accurate reconstruction of the primary cosmic-ray energy and arrival direction.
 
PAO observes cosmic rays as events distributed across the Southern sky, with arrival directions shaped by the distribution of astrophysical sources and by deflections in Galactic and extragalactic magnetic fields~\cite{auger_anisotropy_4EeV_2018, auger_anisotropy_8EeV_2017}.
The observatory provides one of the most important datasets for investigating the origin, composition, and anisotropy of UHECRs~\cite{2015172}. The dataset used in this work~\cite{auger_32_EeV} comprises $2635$ events with reconstructed energies above $32~\mathrm{EeV}$, recorded between 1 January 2004 and 31 December 2020. Each event is described by nine fields: the observation epoch (year, day of the year, and UTC timestamp), the local arrival geometry (zenith and azimuth angles), the celestial arrival direction (right ascension and declination), the reconstructed energy, and the integrated exposure accumulated by the Observatory up to the time of the event.

Events are classified according to their zenith angle into two samples: vertical events with $\theta < 60^\circ$ and inclined events with $60^\circ \leq \theta \leq 80^\circ$. The two samples contain $2040$ and $595$ events, respectively. Vertical and inclined showers are reconstructed with independent procedures, but their relative exposures are consistent with the observed ratio of the two event types, so that the two samples can be
combined into a single dataset. However, for the highest energy UHECRs, a deficit of inclined events becomes apparent. This may reflect limited statistics in this energy range or residual systematic effects in the reconstruction of highly inclined showers.

\begin{figure}[htpb]
    \centering
    \includegraphics[width=\linewidth]{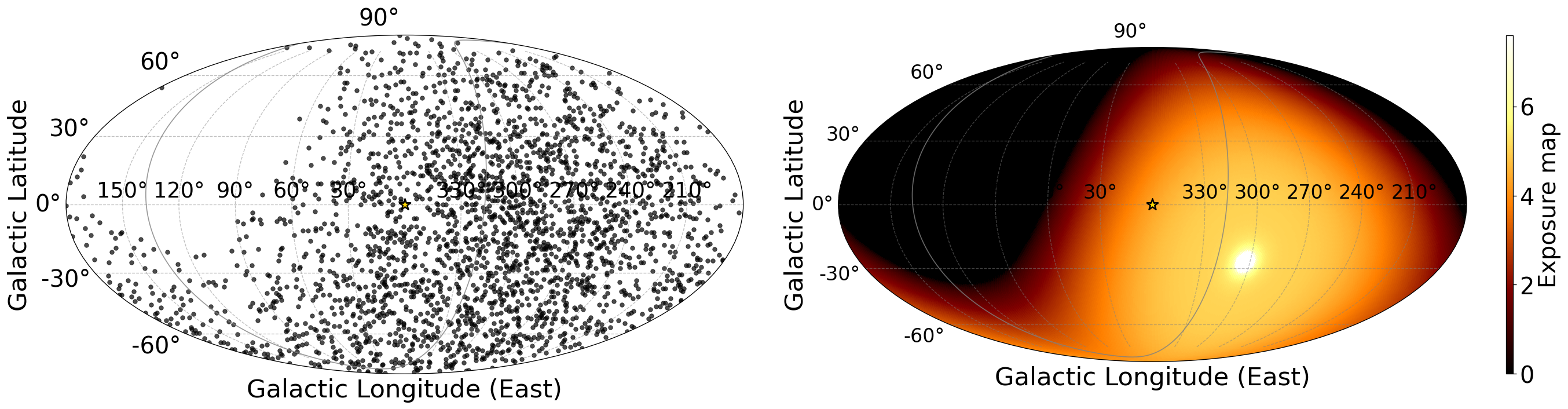}
    \caption{The PAO dataset's arrival-direction count map (left) and the corresponding directional exposure map (right), shown in Galactic coordinates. The yellow star marks the Galactic Center (GC).}
    \label{fig:Auger_data}
\end{figure}

As PAO is located at a latitude of $-35.2^\circ$, its exposure is non-uniform over the sky and vanishes toward northern declinations. The observed distribution of arrival directions therefore carries a strong declination-dependent modulation of purely instrumental origin, which must be accounted for when comparing the data with model predictions. Figure~\ref{fig:Auger_data} shows the PAO dataset's arrival direction map (left panel) and directional exposure map (right panel) in Galactic coordinates. 
As expected, the arrival direction map exhibits a strong declination-dependent asymmetry arising from the non-uniform exposure of the observatory, which is reflected in the exposure map peaking near the detector latitude ($\delta \approx -35.2^\circ$) and decreasing toward higher northern declinations.

PAO has searched for evidence of ultra-relativistic MMs in its UHECR data~\cite{PierreAuger:2016imq}. In that analysis, it was assumed that MMs interact with matter exclusively through electromagnetic energy losses.
Consequently, an MM would produce a signal in the PAO SD and FD that is markedly different from that of a standard UHECR. When an ordinary cosmic ray enters the atmosphere, it initiates electromagnetic and hadronic cascades that are progressively absorbed within a few interaction lengths from the point of the first interaction. By contrast, if an MM interacts exclusively through electromagnetic processes, it would continuously generate secondary showers and radiation throughout its entire passage through the atmosphere. Under this assumption, no MM candidates have been found. However, only a subset of MM models predicts purely electromagnetic interactions. More generally, hadronic interactions also contribute significantly to the energy loss, potentially leading to a substantially different atmospheric signal that may not be distinguishable from that produced by conventional UHECRs. In the remainder of this section, we therefore investigate the constraints on the MM flux that can be derived from the PAO dataset presented in~\cite{auger_32_EeV} without making assumptions about the specific mechanisms responsible for energy loss in the atmosphere. Instead, our analysis relies on the comparison of  the PAO data to the arrival-direction and final kinetic-energy distributions of the MM simulations described in the previous section.

\subsection{Likelihood analysis}

To evaluate the possibility that MMs could contribute to the observed UHECR flux, a series of likelihood-ratio tests is performed. In the first approach, the methodology employed by PAO is closely followed~\cite{auger_32_EeV}, with a profile-likelihood used for the upper-limit construction on the MM flux. Namely, the correlation of UHECR arrival directions with the simulated MM flux patterns is evaluated against isotropy using a likelihood-ratio analysis. The model as a function of direction $u$ is computed in equal-area bins on the sphere using HEALPix v3.70 \cite{Górski_2005} with the parameter  nSide = 64, which corresponds to a pixel size on the order of the angular resolution of the Observatory. The null hypothesis corresponds to the isotropic flux distribution. Accounting for the directional exposure of the array, $\omega(u)$, the isotropic model for the UHECR count density is:
\begin{equation}
\label{eq:n0_dir}
n_0(u) = \frac{\omega(u)}{\sum_i \omega(u_i)},
\end{equation}
where $n_0$ is the UHECR count density under the null hypothesis, and the sum is over the HEALPix pixels indexed over $i$. The alternative hypothesis in which the null one is nested is the sum of an isotropic component and a component corresponding to a specific MM simulation case. The latter,  $n_{\rm MM}(u)$, is constructed by weighting each event by the PAO exposure at its arrival direction, since PAO doesn't observe all directions with equal exposure. The resulting distribution is then normalized to obtain the MM model probability distribution. The amplitude of the MM component is a variable signal fraction, $\alpha$. The model for the UHECR count density then reads
\begin{equation}
\label{eq:n_dir}
n(u) = (1 - \alpha) \times  n_0(u) + \alpha \times n_{\rm MM}(u)
\end{equation}
Both MMs and anti-MMs are included in Eq.~\eqref{eq:n_dir}.

\begin{figure}[htpb]
  \centering
  \begin{subfigure}[b]{0.44\textwidth}
    \centering
    \includegraphics[width=\textwidth]{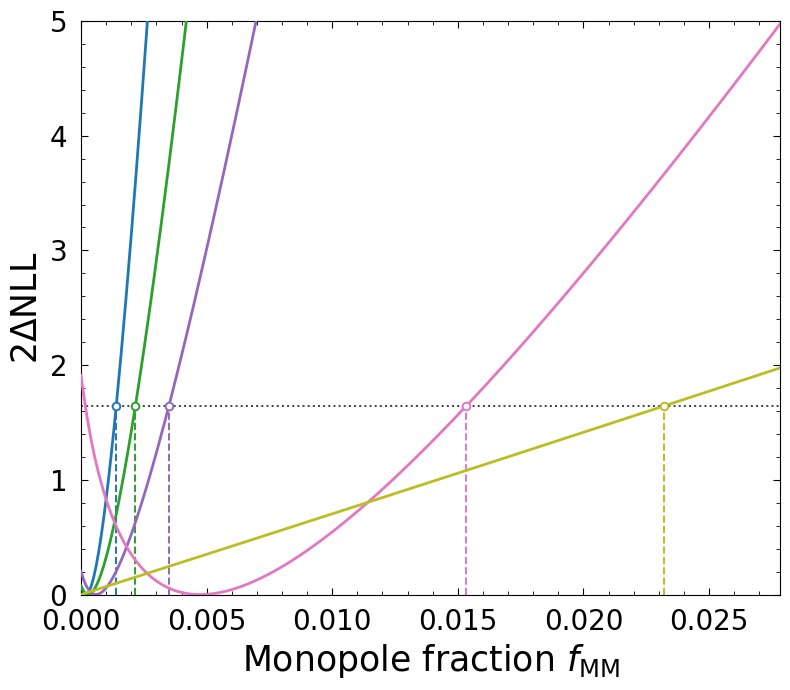}
    \label{fig:nll_fit}
  \end{subfigure}
  \hfill
  \begin{subfigure}[b]{0.52\textwidth}
    \centering
    \includegraphics[width=\textwidth]{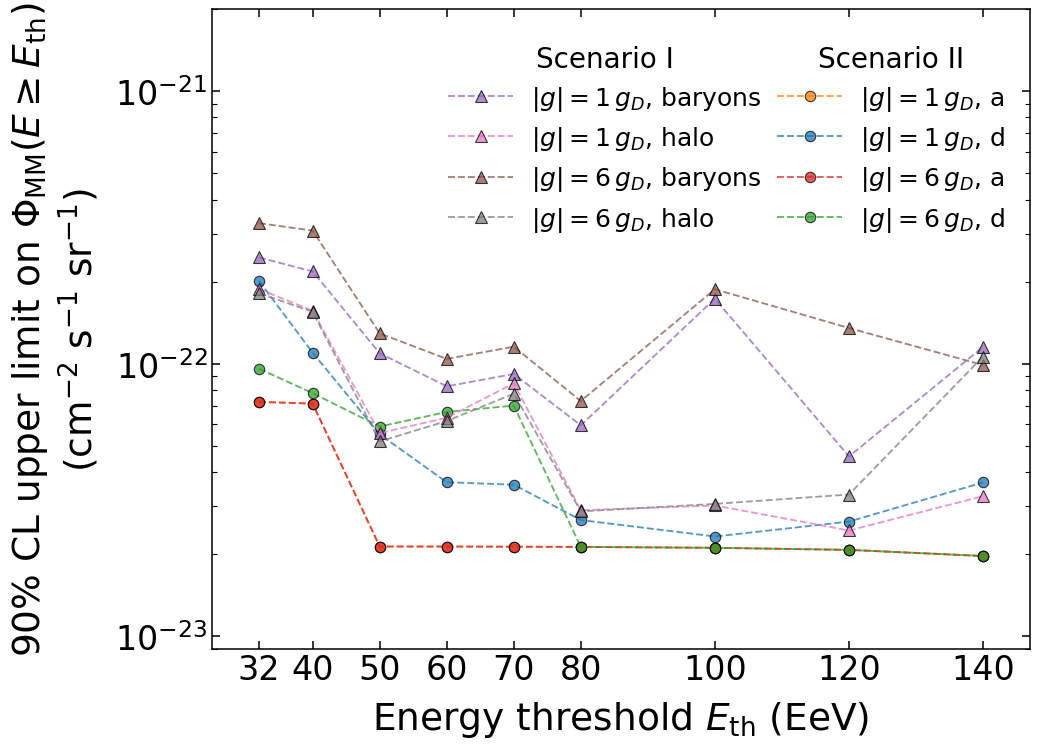}
    \label{fig:flux_vs_energy}
  \end{subfigure}
\caption{\textbf{Left:} Profile likelihoods of the MMs fraction for Auger events above different energy thresholds, $E_{\rm th}$ = 32 (blue), 40 (green), 50 (purple), 70 (pink), 100 EeV (yellow). Scenario IId case with m = 100 TeV/$c^2$ and $|g| = 6 g_D$ is shown. The horizontal dashed line indicates the 90\% confidence-level threshold, while the vertical dashed lines mark the corresponding upper limits of $f_{\rm MM}$. \textbf{Right:} The 90\% CL upper limits on the integral MMs flux at Earth, $\Phi_{\rm MM}(E \geq E_{\rm th})$, as a function of the energy threshold $E_{\rm th}$ for MMs with a mass of 100 TeV/c$^2$. Results are shown for both Scenario I (triangles) and Scenario II (circles) models.}
    \label{fig:fit}
\end{figure}

The test statistic is defined as $\mathrm{TS} = 2\ln(L/L_0)$, where the $L_0$ and $L$ are the likelihoods of the null and alternative hypotheses, obtained as the product of the corresponding models over all events. Rather than testing the null at $\alpha = 0$ directly, a profile-likelihood approach is adopted. That is, the best-fit signal fraction is found by scanning the 1D parameter space with a fixed step, and the likelihood is profiled relative to the best fit. An example for Scenario IId with $m = 100~\mathrm{TeV/c^2}$ and $|g| = 6g_D$ is shown in Figure~\ref{fig:fit}, where the opposite of TS normalized to the best-fit value ($2 \Delta \mathrm{NLL}$) is presented for visual clarity.
Note that at $\alpha = 0$ the curve may therefore be offset from zero, indicating tension between the null hypothesis and data.

For the majority of the considered simulation parameters and energy thresholds, the test results are statistically compatible with the null hypothesis, i.e., no MM contribution to the PAO data. Only 3 out of 16 tested cases fall outside the 90\% confidence interval, all occurring at relatively high energy thresholds. For example, the null hypothesis is excluded at 91.6\% C.L. for the $E_{\rm th}$ = 70 EeV case shown in the left plot of Figure~\ref{fig:fit}. Such occurrence is expected from the frequentist coverage of a 90\% confidence interval, for which the true value is excluded in approximately 10\% of the cases. The right plot of Figure~\ref{fig:fit} shows the 90\% C.L. upper limits on the MM flux at Earth as a function of the energy threshold for the different simulation scenarios. 

The limits are derived using the profile-likelihood test statistics of 1.64, the one-sided critical value appropriate for a single non-negative parameter~\cite{cowan2011asymptotic}. The upper limits on $\alpha$ are converted to an integral MM flux by multiplying it with the number of detected events above given threshold and dividing with $122000~\mathrm{km^2} \times \mathrm{sr} \times \mathrm{yr}$, the total exposure of the detector~\cite{auger_32_EeV}. 

The above approach follows closely the methodology employed by PAO, but it only considers the arrival direction distributions and does not incorporate the information about the expected energy distribution. Whether a general-purpose observatory like PAO could reliably reconstruct energy of an incoming magnetically charged particle is an open question, especially considering that MMs may be subject to QCD interactions in addition to the QED ones. This ambiguity notwithstanding, we extend the first approach by including the energy variable as another observable. For simplicity, potential correlations between the directional and energy observables are ignored, so the total count densities are expressed as products of the direction-only and energy-only normalized distributions. For the MM hypothesis, the energy distribution comes directly from the simulation, as described earlier. For the null hypothesis, the all-particle Gaisser-Stanev-Tilav fourth generation model (GST-4) is used~\cite{gst4}. The model combines data of multiple air shower experiments, including PAO, to construct a joint all-particle energy spectrum spanning from 10$^{14}$ to 10$^{20}$ eV. 

For the joint energy-direction likelihood analysis, no preference for an MM contribution is found for any of the simulated scenarios. Compared with the direction-only analysis, the inclusion of the energy information provides a substantially stronger constraint, mainly because the energy distributions predicted by the MM simulations have only limited overlap with the energy range populated by the PAO data. Consequently, a non-zero MM contribution is more disfavored by the joint likelihood. The resulting 90\% C.L. upper limits on the MM fraction are nearly identical across all cases, corresponding to an upper limit on the integral MM flux at Earth of 
\begin{equation}
    F < 2.1\times10^{-23}~\mathrm{cm^{-2}s^{-1}sr^{-1}} ,
\end{equation}
which is much stronger than previous constraints in the literature (see \cite{Perri_2025} for a recent review).

\section{Conclusion}

In this work, a comprehensive simulation study of the propagation of low-mass MMs through the Galactic magnetic field was conducted. We developed custom modules to accurately treat the dynamics of magnetically charged particles, including energy losses from synchrotron radiation. The key findings are as follows:

\begin{itemize}

\item Relic MMs with low kinetic energy become tightly confined by the coherent Galactic magnetic field. This ``Galactic Magnetic Funnel'' effect forces MMs (and anti-MMs) to arrive at Earth from distinct, highly localized regions of the sky. This anisotropy is a direct consequence of the MMs following the field lines and is robust against changes in the initial spatial and energy distributions, as well as the choice of the UF23 field model;

\item The characteristic arrival kinetic energies of MMs are determined primarily by their magnetic charge, peaking at approximately 10$^3$ EeV for $g=g_{\rm D}$ and 10$^4$ EeV for $g=6g_{\rm D}$. These values are one order of magnitude larger than previous analytical estimates;

\item The strong anisotropy of the arrival directions has significant implications for the existing and next-generation searches for MMs in cosmic rays. A smaller dedicated detector, like the proposed MANDATE experiment~\cite{Mitsou:2026zcf,NextFrontiers}, could potentially outperform a much larger, general-purpose observatory if strategically positioned to cover the expected arrival region;

\item Comparing the simulated distributions with the publicly available PAO UHECR data reveals no statistically significant evidence of a MM component. The likelihood analysis that incorporates both the arrival direction and energy distributions places a 90\% C.L. limit on the integral flux of low-mass MMs at Earth of $\less 2.1\times10^{-23}$ $\mathrm{cm^{-2}s^{-1}sr^{-1}}$. 
\end{itemize}

The results of this work have been derived for MM masses $\lesssim 100~\mathrm{TeV}/c^2$. However, the conclusions are expected to remain valid for substantially heavier MMs, provided that the kinetic energy gained during acceleration in the Galactic magnetic fields is much larger than the MM mass. A dedicated simulation study would be required to verify this. In a forthcoming work, we also plan to extend this simulation framework to non-relativistic MMs, for which the propagation dynamics and resulting anisotropy are expected to differ qualitatively from those in the low-mass regime studied here.

The simulations used to produce the above results omit the turbulent component of the Galactic magnetic field, which is expected to broaden the distributions. However, using the older magnetic field model suggests this to be a mild effect overall, estimated to increase the width of the envelop angular distributions by only few degrees. The effect could be rigorously taken into account with the next update of the field models including the turbulent component, and it is left for future studies. 

Finally, we assumed that MMs produce air-shower profiles and reconstructed energies similar to the ordinary UHECRs, differently from the assumption made in the PAO's MM search, where only electromagnetic interactions were considered. In reality, the MM's atmospheric interactions may differ substantially from either of the two assumptions, shifting the true limits accordingly. 
The uncertainty associated to the adopted model of the MM air-shower profile strengthens one of the main conclusions of this paper: a small experiment optimized for detection of magnetic charge may provide advantages and complementarity compared to a generic, larger shower array because it does not rely on shower-reconstruction assumptions. 

With next-generation Galactic magnetic field models and a strategically placed dedicated detector, there is a chance that the question of whether the most puzzling UHECRs are MMs will be settled conclusively.

\acknowledgments
D.P. was partially supported by the National Science Centre, Poland, under research grant no. 2020/38/E/ST2/00243. We thank Cole Christie for creating the monopole plug-in and Rafael Alves Batista for helpful discussions about CRPropa.  We thank Michele Doro for discussions and comments on this work.

\section*{Data availability}

The UHECR propagation simulations were performed using the publicly available software package CRPropa 3.2.1~\cite{crpropa_2022}. MM propagation was implemented using an external CRPropa plugin developed specifically for particles with magnetic charge. The original plugin is available at \url{https://github.com/chchristie/monopole/tree/main}. All simulations for this work were performed with an updated version of the plugin that fixes a few small issues and adds new features. We plan to merge these updates into the official repository; in the meantime, the updated version is available upon reasonable request.

\section*{AI Use Statement}
The authors used an AI-based language tool to assist with proofreading and editing the manuscript. No AI tools were used for data generation, analysis, or scientific decision-making. The authors take full responsibility for the integrity and accuracy of the work.

\bibliographystyle{ieeetr} 
\bibliography{main}

\newpage

\appendix
\counterwithin{figure}{section}

\section{Effects of the turbulent component of the Galactic magnetic field}
\label{app:turbulent}

\begin{figure}[htpb]
    \centering
    \includegraphics[width=0.7\textwidth]{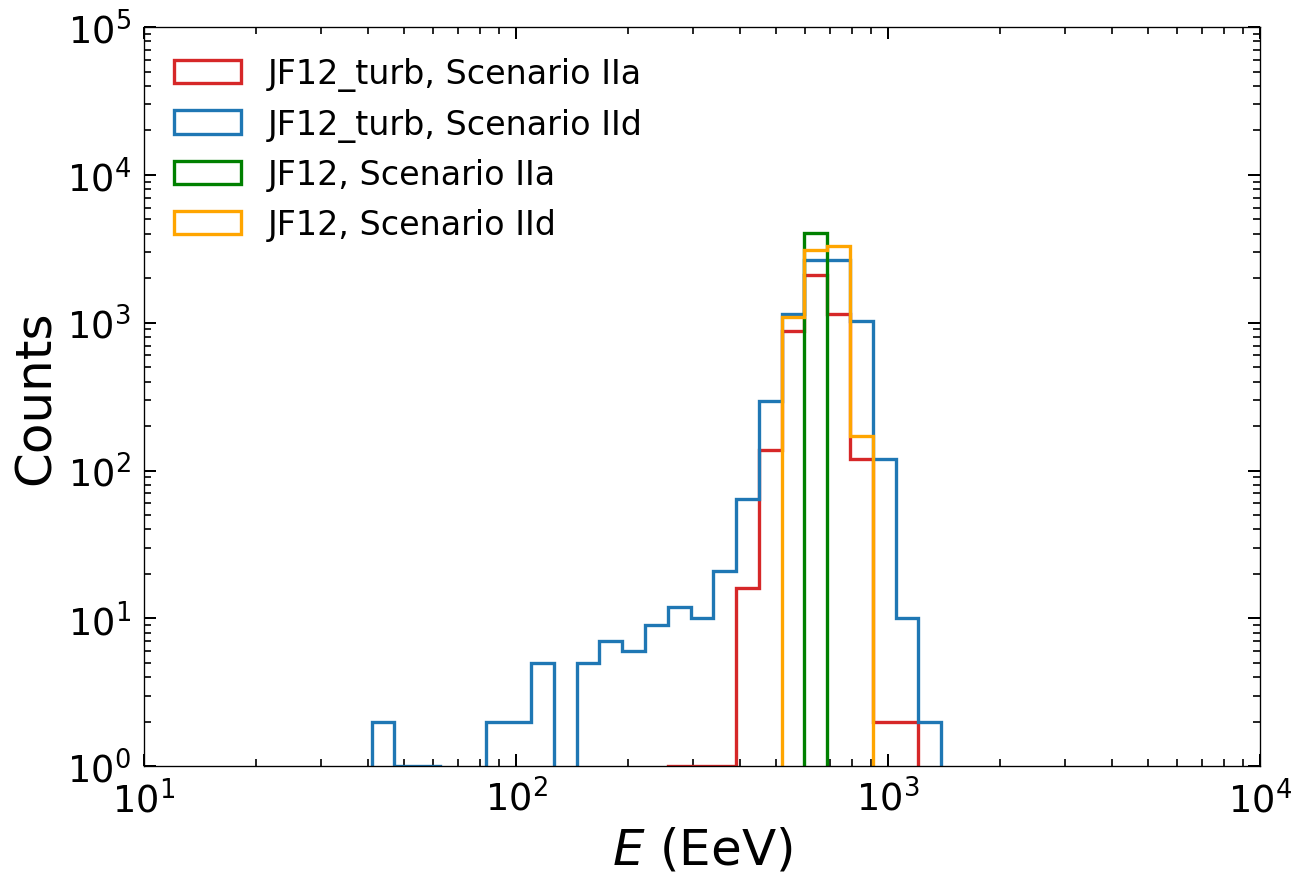}
    \caption{Arrival kinetic-energy distributions of MMs with a mass of 1 TeV/c$^2$ and $g=g_{\rm D}$ propagated in the JF12 field model, with (JF12\_turb) and without (JF12) the turbulent component. Results are shown for the Scenario IIa and IId. The comparison illustrates the effect of turbulent component of the Galactic magnetic field on the MM arrival energy distribution.}
    \label{fig:JF12_energy}
\end{figure}

\begin{figure}[htpb]
    \centering
    \includegraphics[width=\textwidth]{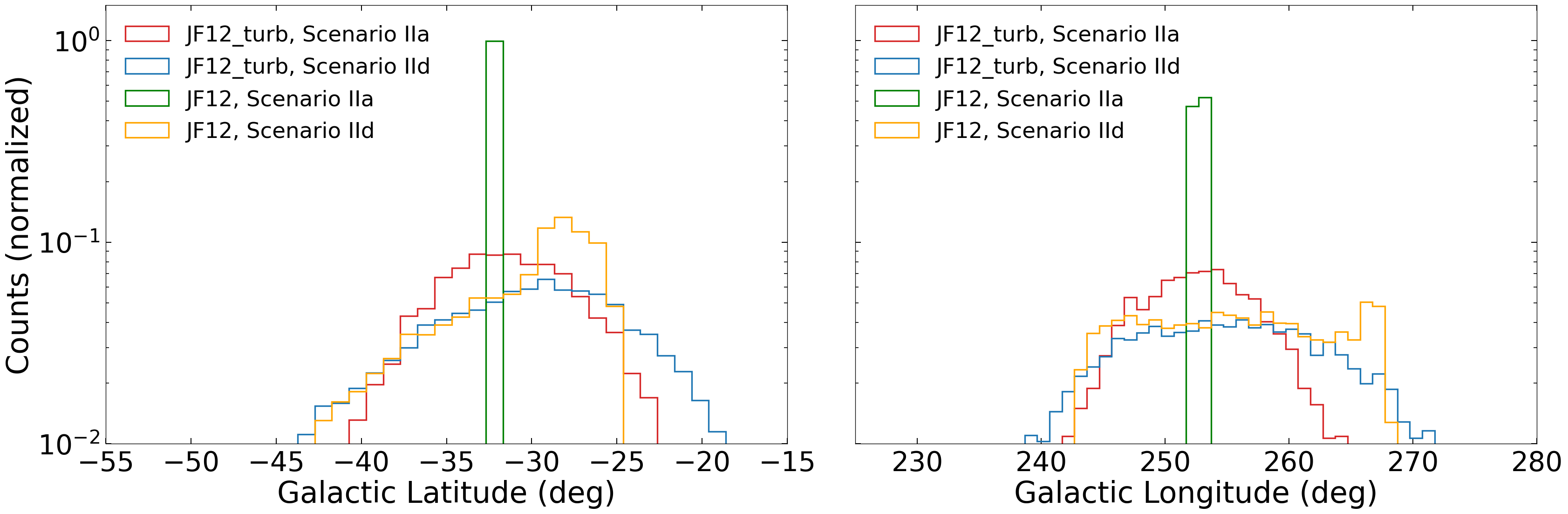}
    \caption{Arrival angular distributions of MMs with a mass of 1 TeV/c$^2$ and $g=g_{\rm D}$ propagated in the JF12 field model, with (JF12\_turb) and without (JF12) the turbulent component. Results are shown for the Scenario IIa and IId. The comparison illustrates the effect of turbulent component of the Galactic magnetic field on the MM arrival angular distribution.}
    \label{fig:JF12_angular}
\end{figure}

\newpage

\section{Results for the UF23 expX magnetic field model}
\label{app:expX}
\begin{figure}[htpb]
  \centering
  \begin{subfigure}[b]{0.49\textwidth}
    \centering
    \includegraphics[width=\textwidth]{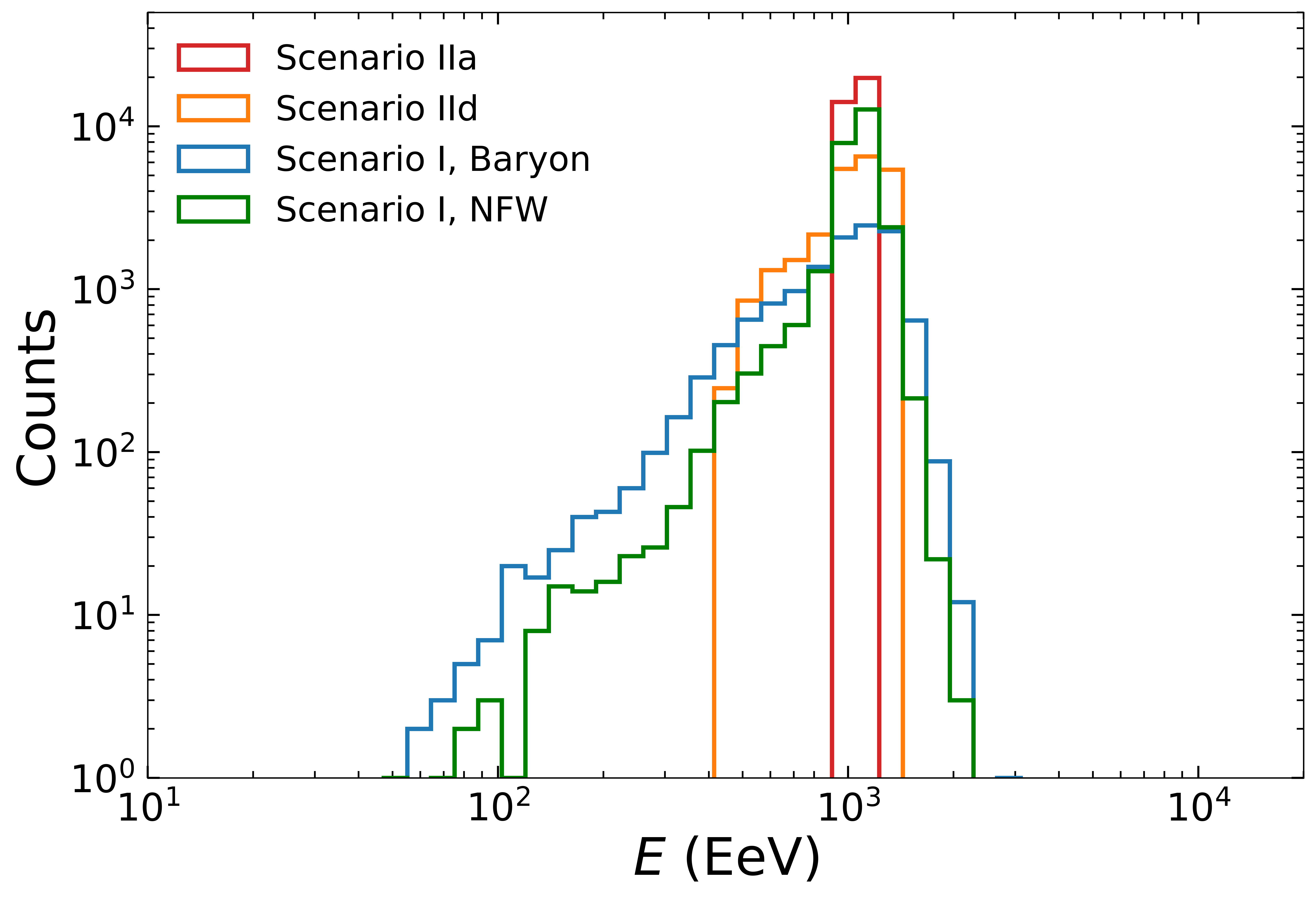}
    \caption{$g=1g_{\rm D}$}
    \label{fig:ek_expX_1gD}
  \end{subfigure}
  \hfill
  \begin{subfigure}[b]{0.49\textwidth}
    \centering
    \includegraphics[width=\textwidth]{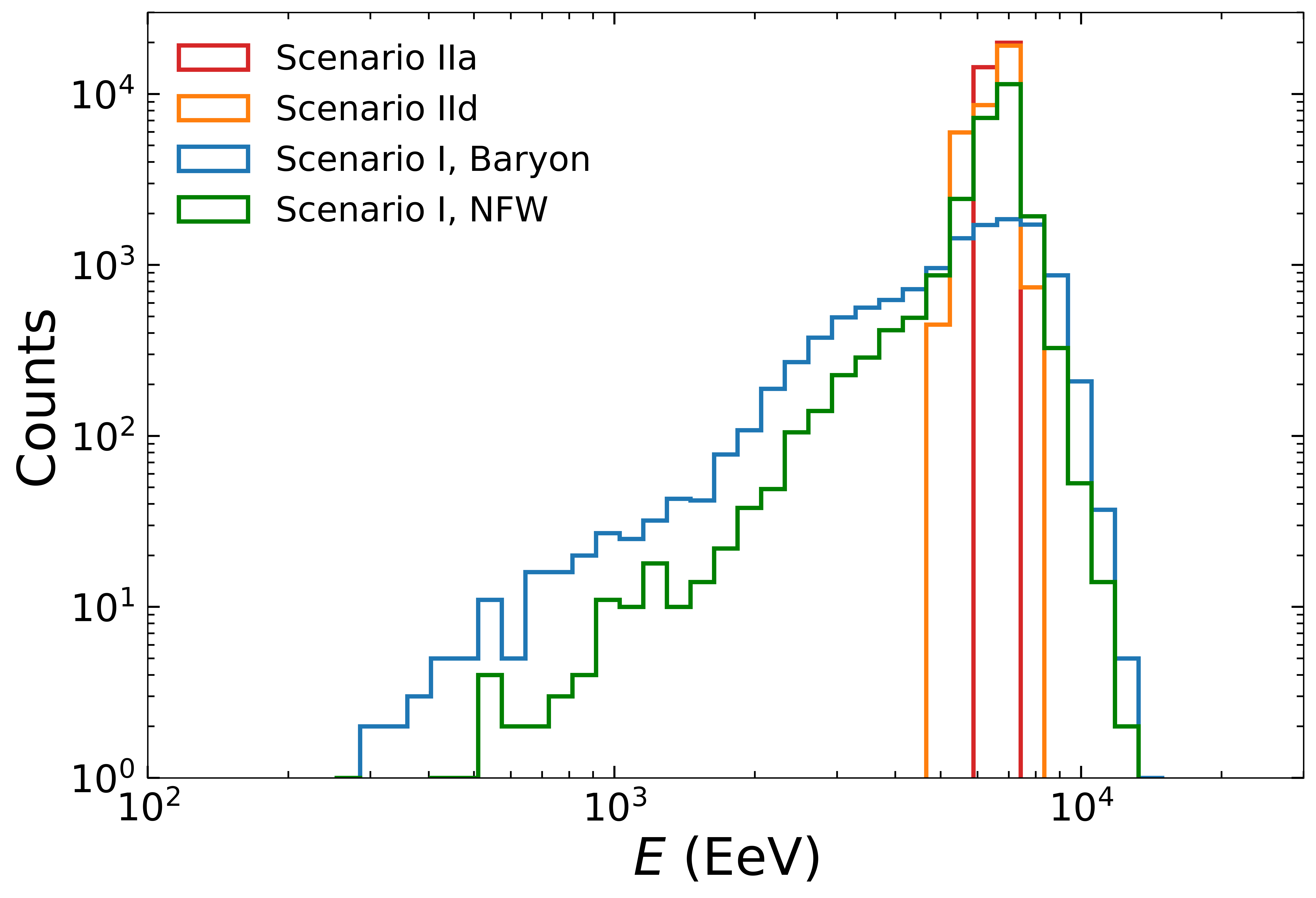}
    \caption{$g=6g_{\rm D}$}
    \label{fig:ek_expX_6gD}
  \end{subfigure}

  \caption{Same as Fig.~\ref{fig:ek_base}, but for the UF23 expX model.}
  \label{fig:ek_expX}
\end{figure}

\begin{figure}[htpb]
  \centering
  \begin{subfigure}[b]{\textwidth}
    \centering
    \includegraphics[width=\textwidth]{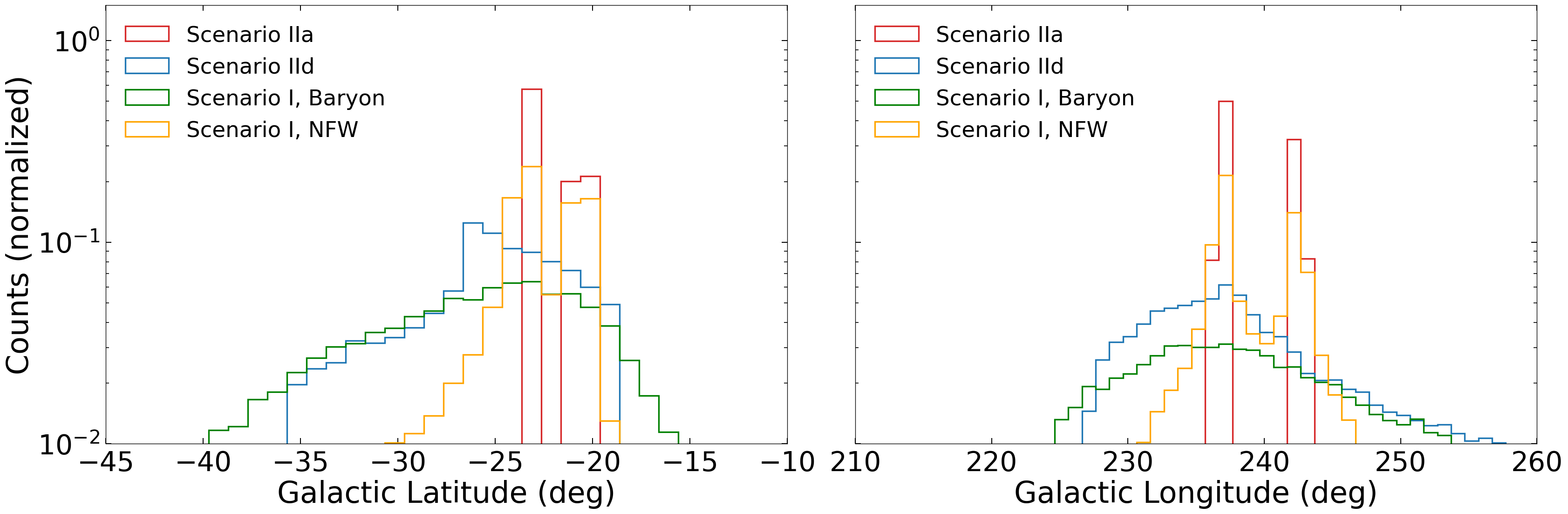}
    \caption{$g=g_{\rm D}$}
    \label{fig:angles_expX_1gD}
  \end{subfigure}
  \vspace{0.5cm}
  \begin{subfigure}[b]{\textwidth}
    \centering
    \includegraphics[width=\textwidth]{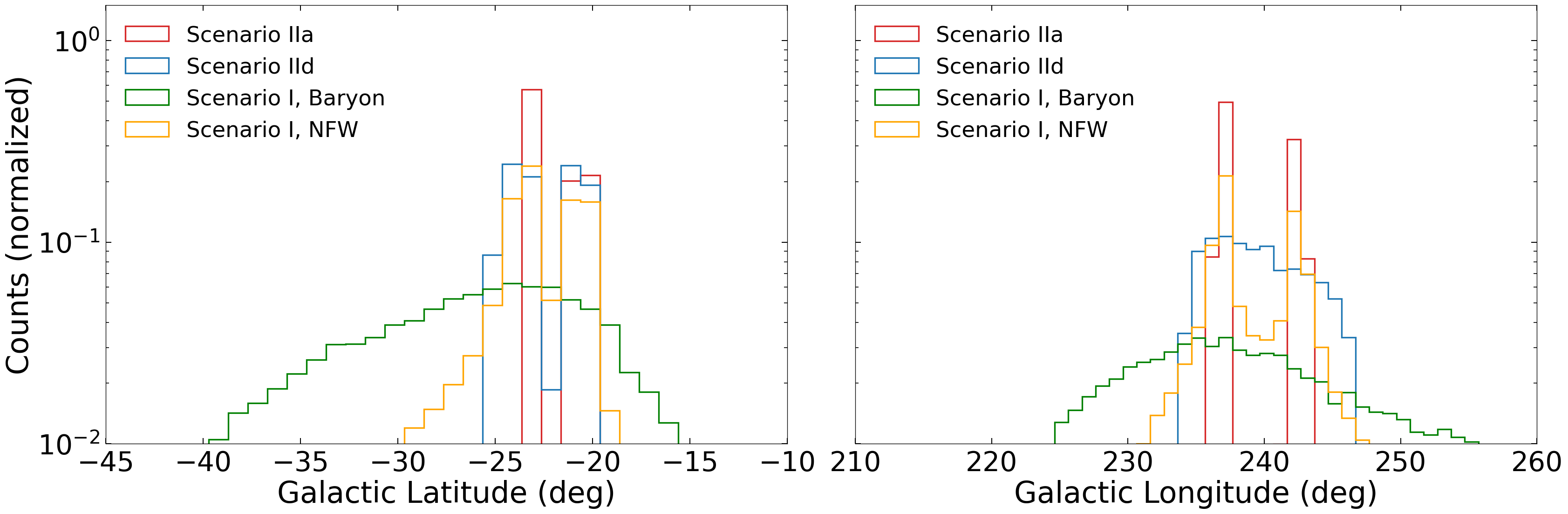}
    \caption{$g=6g_{\rm D}$}
    \label{fig:angles_expX_6gD}
  \end{subfigure}

  \caption{Same as Fig.~\ref{fig:angles_base}, but for the UF23 expX model.}
  \label{fig:angles_expX}
\end{figure}

\captionsetup[subfigure]{labelformat=empty}
\begin{figure}[htpb]
  \centering
  \begin{subfigure}[b]{\textwidth}
    \centering
    \includegraphics[width=\textwidth]{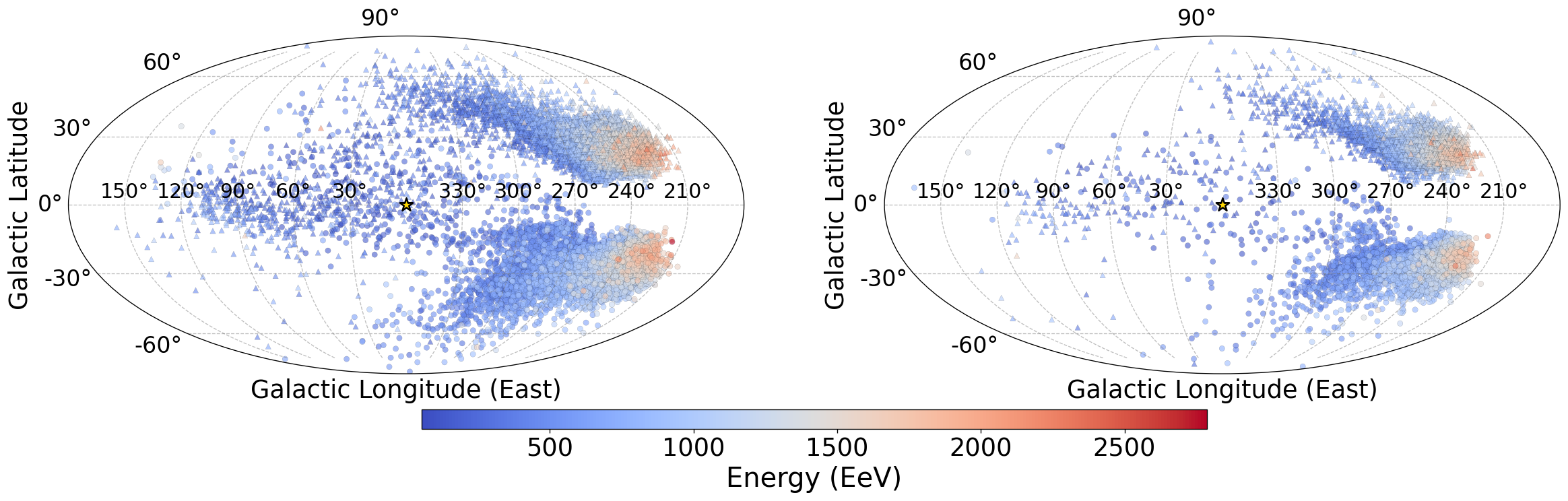}
    \caption{ a) Baryon, $g=1g_{\rm D}. $\hspace{5cm}   b) NFW, $g=1g_{\rm D}$.}
    \label{fig:II_expX_1gD}
  \end{subfigure}
  \vspace{0.5cm}
  \begin{subfigure}[b]{\textwidth}
    \centering
    \includegraphics[width=\textwidth]{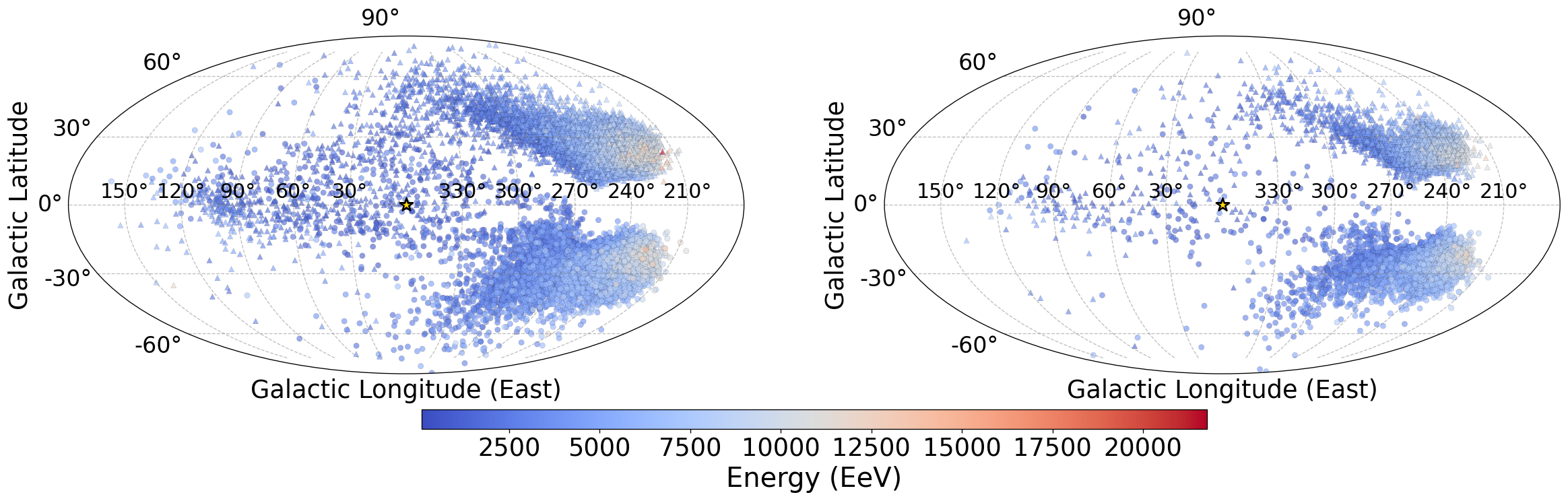}
    \caption{ c) Baryon, $g=6g_{\rm D}$. \hspace{5cm}  d) NFW, $g=6g_{\rm D}$.}
    \label{fig:II_expX_6gD}
  \end{subfigure}
\caption{Same as Figure~\ref{fig:scII_skymaps}, but for the UF23 expX Galactic magnetic-field model.}
    \label{fig:scII_expX_skymaps}
\end{figure}

\captionsetup[subfigure]{labelformat=empty}
\begin{figure}[htpb]
  \centering

  \begin{subfigure}[b]{\textwidth}
    \centering
    \includegraphics[width=\textwidth]{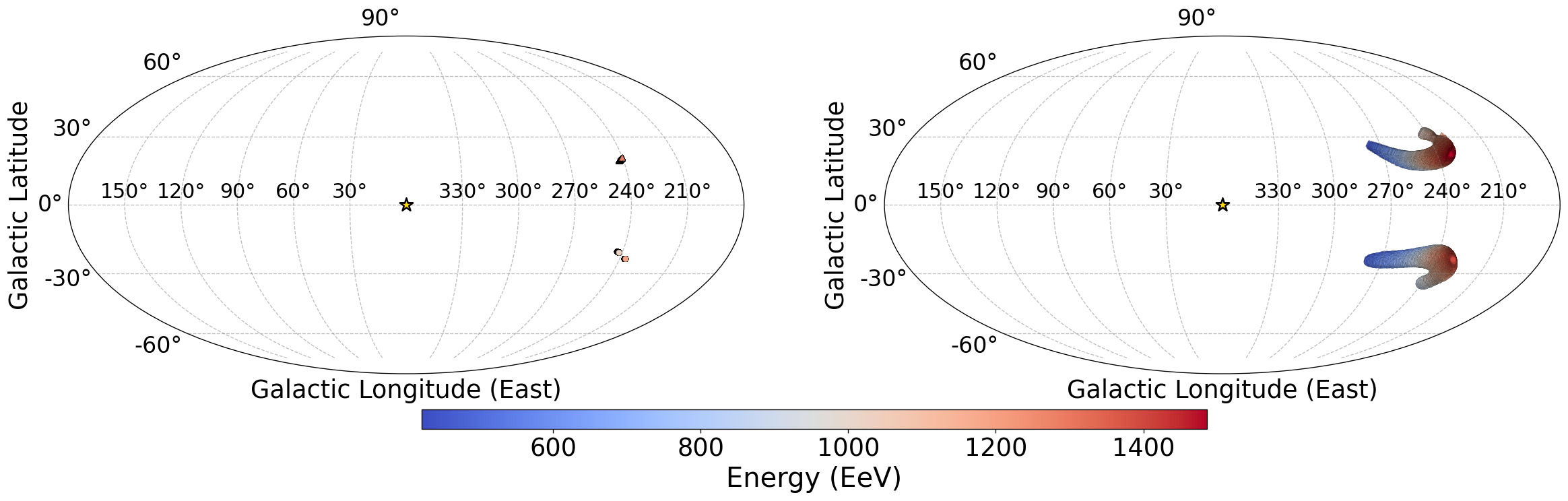}
    \caption{ a) Scenario IIa, $g=1g_{\rm D}$\hspace{3cm}
              b) Scenario IId, $g=1g_{\rm D}$}
    \label{fig:III_1gD_expX}
  \end{subfigure}

  \vspace{0.5cm}

  \begin{subfigure}[b]{\textwidth}
    \centering
    \includegraphics[width=\textwidth]{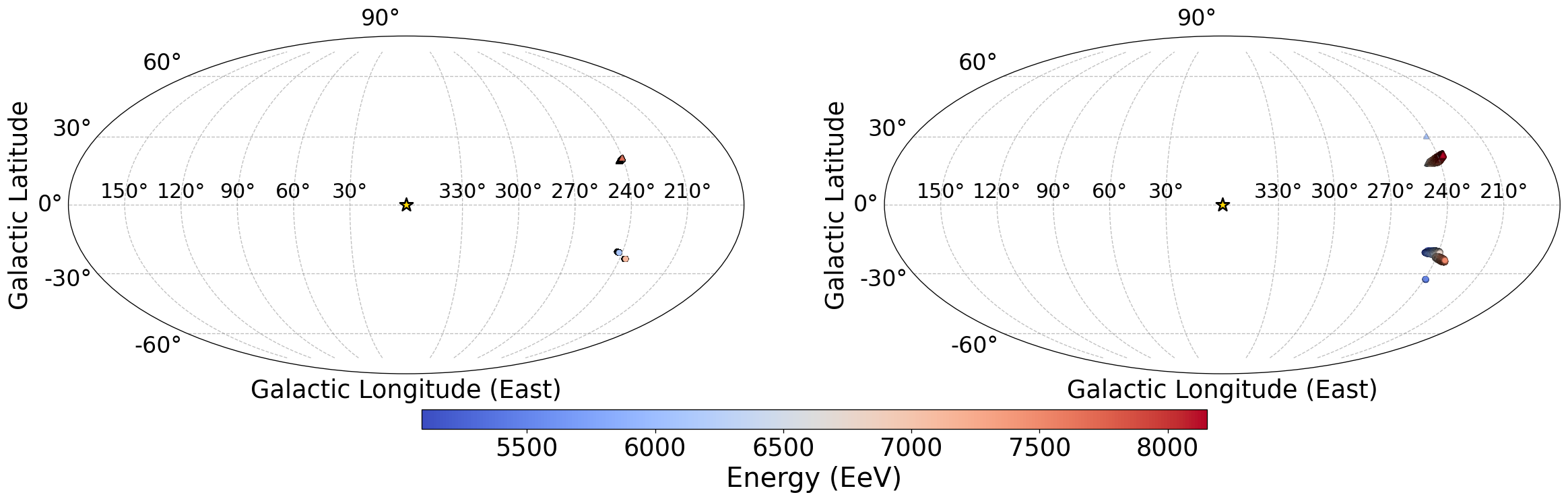}
    \caption{ c) Scenario IIa, $g=6g_{\rm D}$. \hspace{3cm}
              d) Scenario IId, $g=6g_{\rm D}$.}
    \label{fig:III_6gD_expX}
  \end{subfigure}
  \caption{Same as Figure~\ref{fig:skymap_III}, but for the UF23 expX Galactic magnetic-field model.}
  \label{fig:skymap_III_expX}
\end{figure}

\clearpage

\section{Results for Scenario IIe}
\label{app:IIIf}

\begin{figure}[htpb]
    \centering
    \includegraphics[width=0.95\textwidth]{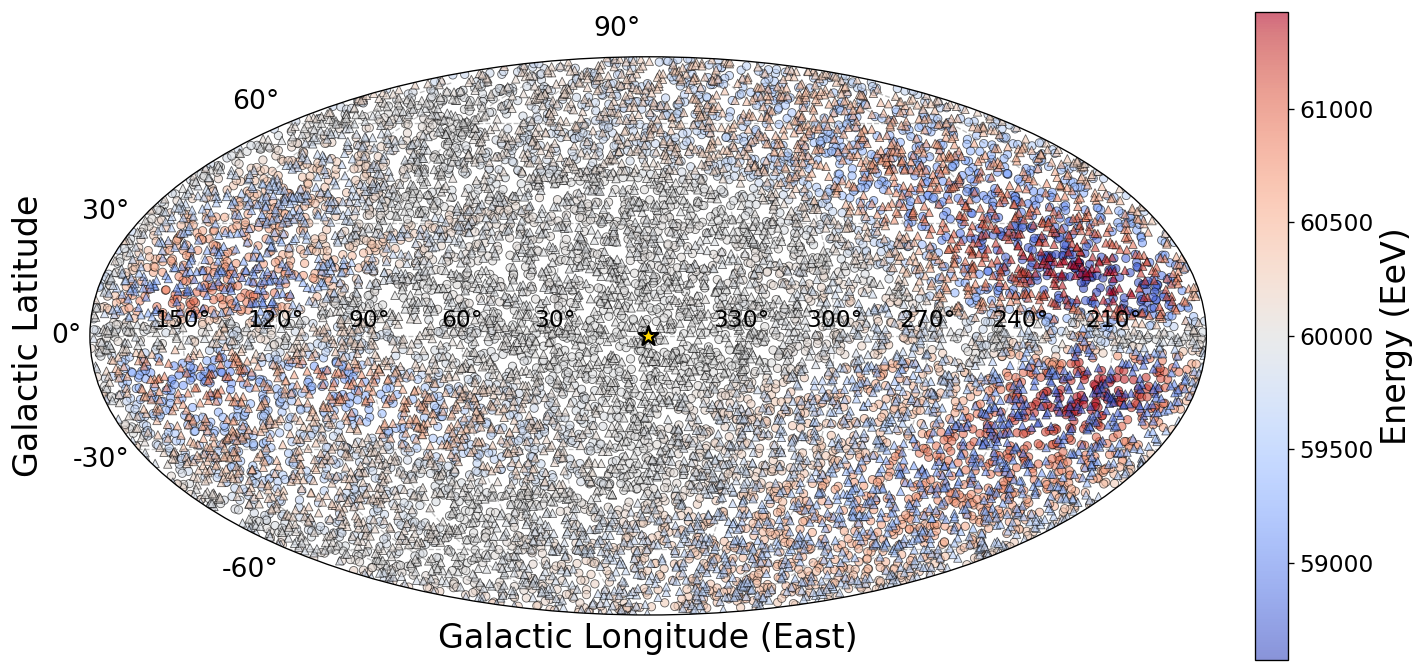}
    \caption{Skymap of the arrival directions of relic MMs (circles) and anti-MMs (triangles) reaching Earth from an initially isotropic distribution for the extreme case of Scenario II, with an initial kinetic energy of $6\times10^{4}~\mathrm{EeV}$, in the presence of a strong intergalactic magnetic field. The MM mass is fixed to $m=1~\mathrm{TeV}/c^2$, and two choices of the magnetic charge, $g=1g_{\rm D}$ and $g=-1g_{\rm D}$, are considered. The UF23 base model is adopted for the Galactic magnetic field. The yellow star marks the Galactic Center (GC). The color scale shows only tiny deviations from $6\times10^{4}~\mathrm{EeV}$, indicating that the energy gain inside the Galaxy is negligible compared with the initial MM kinetic energy.}
    \label{fig:IG1_extreme}
\end{figure}

\end{document}